\documentclass[twocolumn,showpacs,amsmath,amssymb,superscriptaddress]{revtex4}

\usepackage{epsfig}
\usepackage{graphicx}

\newcommand{\abs}[1]{\left|{#1}\right|}
\newcommand\intd{\ensuremath{\mathrm{d}}}
\newcommand\vect[1]{\ensuremath{\mathbf{#1}}}
\newcommand\expect[1]{\left\langle{#1}\right\rangle}
\newcommand\vgrad{\ensuremath{\vect{\nabla}}}
\newcommand\rrho{\ensuremath{\hat\varrho}}
\newcommand\aop{\ensuremath{\Hat{a}}}
\newcommand\sop{\ensuremath{\Hat{\sigma}}}
\newcommand\Fopk{\ensuremath{\hat{\vect{F}}_k}}
\newcommand\bkm{\ensuremath{{\boldsymbol \beta}_{km}}}
\newcommand\dkm{\ensuremath{{\mathcal D}_{km}}}
\newcommand\xk{\ensuremath{\vect{x}_k}}
\newcommand\xl{\ensuremath{\vect{x}_l}}
\newcommand\dd{\ensuremath{\mathcal{D}}}

\begin{document}

\title{Correlated motion of two atoms trapped in a single mode cavity field}

%\author{J\'anos K.\ Asb\'oth$^{1,2}$}
%\author{Peter Domokos$^{1}$}
%\author{Helmut Ritsch$^{2}$}
%%
%
%\affiliation{$^1$ Research Institute of Solid State Physics and Optics,
%Hungarian Academy of Sciences\\
%H-1525 Budapest P.O. Box 49, Hungary\\
%$^2$ Institute of Theoretical Physics, University of Innsbruck, 
%Technikerstrasse 25, A-6020 Innsbruck, Austria
%}
\author{J\'anos K.\ Asb\'oth}
\affiliation{Research Institute of Solid State Physics and Optics,
Hungarian Academy of Sciences,
H-1525 Budapest P.O. Box 49, Hungary}
\affiliation{Institute of Theoretical Physics, University of Innsbruck, 
Technikerstrasse 25, A-6020 Innsbruck, Austria}
\author{Peter Domokos}
\affiliation{Research Institute of Solid State Physics and Optics,
Hungarian Academy of Sciences,
H-1525 Budapest P.O. Box 49, Hungary}
\author{Helmut Ritsch}
\affiliation{Institute of Theoretical Physics, University of Innsbruck, 
Technikerstrasse 25, A-6020 Innsbruck, Austria}

\begin{abstract}
  We study the motion of two atoms trapped at distant positions in the
  field of a driven standing wave high $Q$ optical resonator.  Even
  without any direct atom-atom interaction the atoms are coupled
  through their position dependent influence on the intracavity field.
  For sufficiently good trapping and low cavity losses the atomic
  motion becomes significantly correlated and the two particles
  oscillate in their wells preferentially with a $90^{\circ}$ relative
  phase shift.  The onset of correlations seriously limits cavity
  cooling efficiency, raising the achievable temperature to the
  Doppler limit.  The physical origin of the correlation can be traced
  back to a cavity mediated cross-friction, i.e.~a friction force on
  one particle depending on the velocity of the second particle.
  Choosing appropriate operating conditions allows for engineering
  these long range correlations. In addition this cross-friction
  effect can provide a basis for sympathetic cooling of distant trapped
  clouds.
\end{abstract}

\pacs{32.80.Pj, 33.80.Ps, 42.50.Vk}

\maketitle
\section{Introduction}
\label{sect:intro}

It is a well established fact, both theoretically and experimentally,
that light forces on atoms are substantially modified within resonant
optical cavities
\cite{domokos03,vuletic01,vanenk01,munstermann00,kruse03,doherty97,mossberg91,rempe04}.
Possible experimental realizations range from single atoms or ions
\cite{mundt02,guthohrlein02} in microscopic super cavities to several
thousand \cite{vuletic03a,chapman03} or up to a million atoms
\cite{kruse03,zimmermann03} in a high $Q$ ring cavity. Applications of
these systems include possible implementations of quantum information
processing setups \cite{hemmerich99,griessner03}, and controlled
nonclassical light sources \cite{kimble03,kuhn02} as well as new
possibilities for trapping and cooling of atoms and molecules. The
basic physical mechanism in these setups can be traced back to the
backaction of the atoms on the field.  They act as a moving refractive
index and absorber, modifying the intensity and phase of the
intracavity field, which in turn governs their motion.  This coupled
dynamics is at the heart of cavity enhanced trapping and cooling.

It is clear that if a single atom is able to change the field, it will
influence other atoms in the same field irrespective of their
distance.  This introduces long range atom-atom interactions, which
are widely tailorable by suitable choices of cavity geometries and
operating conditions. On one hand these interactions are useful and
can be used to implement bipartite quantum gates \cite{hemmerich99}.
On the other hand they play a decisive role in the scaling properties
of cavity enhanced cooling \cite{horak01,fischer01}.  For perfectly
correlated atoms, the change of refractive index induced by one atom
can be compensated by a second atom, so that the effective atom-field
back reaction can be strongly reduced.  For several atoms in a ring
cavity this effect only allows a weak damping of relative motion,
while the center of mass motion is strongly damped \cite{gangl00b}.
This model is closely related to the so-called CARL laser, where the
kinetic energy of an atomic beam leads to gain into the
counterpropagating mode of a single side pumped ring resonator
\cite{bonifacio97}. New effects were also found in the study of the
coupling of two Bose-condensates in a cavity \cite{jaksch01}.

Several limiting cases for N atoms commonly interacting with a cavity
mode have already been studied. For the case of N strongly trapped
atoms in a standing wave cavity mode, it is possible to derive a set
of coupled equations for the total kinetic and potential energy as
well as the field amplitude \cite{gangl99}, which exhibit collective,
damped oscillations ending in highly correlated steady states.  This
approach, however, does not give much insight into the details of the
individual dynamics and correlations.  In the opposite limit of N
untrapped atoms moving in the cavity field, numerical simulations show
little influence of atom-atom correlations and cooling proceeds
independent of the atom number \cite{horak01} for proper rescaling of
the cavity parameters.
 
Recently an approach for several atoms in a single mode cavity has
been developed, which concentrates on the effect of the $N-1$ other
particles on the cooling properties of a single one \cite{fischer01}.
This, in principle, makes it possible to study the combined optical
potential and friction forces.  It has been recently proposed
theoretically \cite{domokos02b} and confirmed experimentally
\cite{vuletic03}, that if the atoms are pumped directly from the side
(as opposed to pumping the cavity), the buildup of spatial
correlations within a cloud of trapped atoms can lead to superradiant
light scattering and enhanced cooling behavior. The theoretical
results in this case are based on numerical simulations of the
semiclassical equations of motion for a large number $N \gg 1$ of
particles \cite{domokos02b}. This clearly demonstrates collective
effects, but does not give much quantitative insight in the buildup
and role of atom-atom correlations.

The central goal of the present work is to study the basic physical
mechanisms responsible for the motional correlations and to develop
quantitative measures of the established steady state correlation. For
this we restrict ourselves to the simplest nontrivial example, namely
two atoms strongly coupled to a single standing wave field of a
cavity.  The energy loss is compensated by cavity pumping, and large
detuning from the atomic transition is taken to ensure low atomic
saturation.  Moreover, the motion of the atoms is only followed along
the cavity axis. As we will see, this contains most of the essential
physics but still allows us to derive analytical expressions for many
relevant quantities. Most of the analytical results are valid for the
more general $N$-atom case.

The article is organized as follows: after presenting our model and
the approximations used in section II, we analytically discuss the
central physical mechanisms present in section III. In section IV we
quantify the results using numerical simulations, which are then
analyzed in more detail in section V.

\section{The model}
\label{sect:themodel} 

\begin{figure}
\begin{center}
\includegraphics[width=9cm]{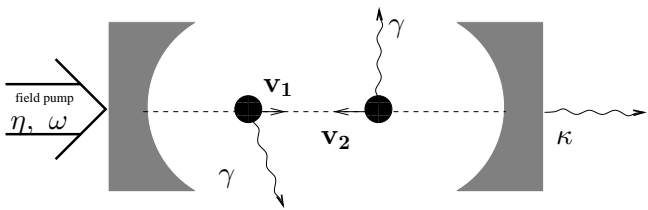}
\caption{\emph{The setup.} 
The motion of two two-level atoms in a far-detuned, high-Q optical cavity is
studied.
The cavity is pumped by a strong laser beam, almost on resonance with
the cavity mode. Atomic motion is discussed along the cavity axis only.
\label{fig:setup} 
}
\end{center}
\end{figure}

Let us start by outlining the system which is shown in
Fig.~\ref{fig:setup}.  We consider $N=2$ two-level atoms with
transition frequency $\omega_A$ strongly coupled to a single mode of a
high-finesse cavity with frequency $\omega_C$.  The system is driven
by a coherent laser field of frequency $\omega$ and amplitude $\eta$
injected into the cavity through one of the mirrors.  The model and
its theoretical treatment follow closely Ref.~\cite{domokos03}.
Coupling to the environment introduces a damping via two channels.
First, the atoms spontaneously emit with a rate of $2\gamma$ into the
vacuum outside the cavity.  Second, the cavity photons decay with rate
$2\,\kappa$ via the output coupler mirror of the cavity.  The atoms
can move freely in the cavity, however, for the sake of simplicity,
their motion is restricted to the cavity axis (dashed line in
Fig.~\ref{fig:setup}).

Applying the standard Born-Markov approximation the dynamics are
governed by a quantum master equation,
\begin{equation}
\label{eq:master}
  \dot\rho = -\frac{i}{\hbar}[\Hat H, \rho] + \cal L \rho.
\end{equation}

Using rotating-wave and dipole approximations the Hamiltonian and the
Liouville operators in an interaction picture read \cite{domokos03}:
\begin{subequations}
\label{eq:H}
\begin{align}
  \Hat H & = \sum_{k=1}^{N} \left[ \frac{\hat{\vect p}_k^2}{2m} -
    \hbar\Delta_A\sop_k^z 
\right] \nonumber\\
  & \quad - \hbar\Delta_c \aop^\dagger \aop -
    i\hbar \eta\left(\aop - \aop^\dagger\right) \; \nonumber
    \\
  & \quad - i\hbar
   \sum_{k=1}^N \left( g(\hat{\vect x}_k) \sop_k^\dagger \aop -
    g^*(\hat{\vect x}_k) \aop^\dagger \sop_k \right)
  \\
  {\cal L}\rrho & = \kappa\left(2 \aop\rrho \aop^\dagger - 
    \aop^\dagger \aop \rrho - \rrho \aop^\dagger \aop \right) \nonumber \\ 
  & \quad + \gamma \sum_{k=1}^{N}
  \Bigl( 2 \int d^2{\vect u}\; N({\vect u})\sop_k e^{-i k_A {\vect u} \hat{\vect
        x}_k} \rrho e^{i k_A {\vect u} \hat{\vect x}_k} \sop_k^\dagger 
    \nonumber \\ 
   & \quad - \sop_k^\dagger \sop_k \rrho - \rrho \sop_k^\dagger
    \sop_k \Bigr) \, .
\end{align}
\end{subequations}
The atomic and cavity field detunings are defined as
$\Delta_A=\omega-\omega_A$ and $\Delta_C=\omega-\omega_C$,
respectively.  The annihilation and creation operators of the cavity
field are $\aop$ and $\aop^\dagger$, while $\sop_k$ and
$\sop_k^\dagger$ are the lowering and raising operators of the $k$-th
atom. The Hamiltonian consists of the motional and internal energy of
the atoms, the self-energy of the cavity field, the classical (laser)
pump, and the Jaynes--Cummings-type interaction between the atoms and
the field. The position dependence of the coupling constant is due to
the spatial structure of the field mode: $g(\xk) = g_0 f(\xk)$, where
in our standing-wave cavity $f(\xk)=\cos(k_C x_k)$, with $k_C = 2 \pi
/ \lambda$ being the cavity wavenumber.  The Liouvillean operator
includes the effect of cavity losses and of the spontaneous emissions
on the combined atom-cavity field density operator $\rrho$.  This
latter is given by the last term, where the integral goes over the
directions of photons spontaneously emitted by the atomic dipole, 
having expected wavenumber $k_A=\omega_A/c$ and angular
distribution $N(\vect{u})$.

We consider cold atoms but with a temperature well above the recoil
limit $k_B T_{rec}=\hbar k_A^2/(2M)$, where $M$ is the mass of one atom.  In
this limit the atomic coherence length is smaller than the optical
wavelength and the position and momentum of the atoms can be replaced
by their expectation values and treated as classical variables. We
still keep the quantum nature of the internal variables $\aop$ and
$\sop_k$.  Moreover, if the atoms move much less than a wavelength
during the equilibration time of the internal variables
\begin{equation}
  \label{eq:adiabaticity}
  v \ll \lambda \kappa, \lambda \gamma,
\end{equation}
we can adiabatically separate the ''fast'' internal from the ''slow''
external dynamics as in standard laser cooling models \cite{domokos02}. 

\subsection{The internal dynamics}
For given positions of the atoms $\xk$ the internal atomic dynamics
can be rewritten in the form of quantum Langevin equations.  For low
saturation, i.e.~when $\expect{\sop^\dagger\sop}\ll 1$, we
can approximate the operator $\hat \sigma_z$ by $-1/2$ (this is called
bosonization of the atomic operators).  The resulting
Heisenberg--Langevin equations then reduce to the following set of
coupled linear differential equations:
\begin{subequations}
\label{eq:bosonic_Langevin}
\begin{align}
  \frac{\intd}{\intd t}{\aop} &= (i\Delta_C - \kappa) a + 
  \sum_k g^*({\vect x}_k) \sop_k + \eta + \Hat\xi \\
 \frac{\intd}{\intd t} {\sop}_k &= (i \Delta_A - \gamma) \sop_k - 
  g({\vect x}_k)\aop 
+ \Hat\zeta_k .
\end{align} 
\end{subequations}
The noise operators $\hat \xi$ and $\hat \zeta_k$ appear as a result
of the coupling to the external vacuum through the cavity mirrors and
through spontaneous emission.  They contain the annihilation operators
of the external vacuum modes, and therefore give $0$ when acting on the
environment's state.  Their second-order correlation functions are as
follows:
\begin{subequations}
\label{eq:zaj}
\begin{align}
  \langle \Hat\xi(t_1) \Hat\xi^+(t_2) \rangle &= 2 \kappa\,
  \delta(t_1-t_2),\\
  \langle \Hat\zeta_k(t_1) \Hat\zeta^+_m(t_2) \rangle &= 2 \gamma\,
  \delta_{km} \delta(t_1-t_2), 
\end{align} 
\end{subequations}
while all other correlations vanish. 

The steady state expectation values of the internal variables $\hat a$ and 
$\hat \sigma_k$ obtained from the Heisenberg--Langevin equations 
(\ref{eq:bosonic_Langevin}) then read:
\begin{subequations}
  \label{eq:ss_mean_a_and_sigma}  
  \begin{align}
 \expect{\aop} &= 
\eta \frac{\gamma - i \Delta_A}{\dd'}
\\ 
\expect{\sop_l} &= 
-\eta 
 \frac{g(\xl)}{\dd'}.
  \end{align}
\end{subequations}
Here $\dd'$ is the reduced determinant of the Bloch
matrix,
\begin{equation}
   \label{d_prime_def}
\dd' = (i\Delta_C -\kappa)(i\Delta_A-\gamma) + \sum_l g(\vect{x}_l)^2.
\end{equation}
Since the factor $1/\dd'$ appears in both expectation values and in
later formulae as well, it is worthwhile to rewrite it to reveal its
resonance structure with respect to the cavity detuning:
\begin{align}
   \label{eq:rezonancia_D}
\frac{1}{\dd'}
&= \frac{1}{(i\Delta_A-\gamma)}\, 
\frac{1}{i(\Delta_C-U)-(\kappa + \Gamma)},\\
   \label{eq:U0_def}
& \mathrm{where} \qquad U 
= \frac{\Delta_A \sum_l g^2(\xl)}{\Delta_A^2 + \gamma^2} 
= U_0 \sum_l f^2(\xl),\\
   \label{eq:Gamma0_def}
& \mathrm{and} \qquad \quad \Gamma =
\frac{\gamma \sum_l g^2(\xl)}{\Delta_A^2 + \gamma^2}
= \Gamma_0 \sum_l f^2(\xl),
\end{align}
and we used $g(x) = g_0 f(x)$. It is clearly seen that each atom
broadens the resonance at most by $\Gamma_0$ and displaces it by
$U_0$.

\subsection{The external dynamics}
The motion of the atoms is governed by the force operator,
\begin{equation}
\begin{split}
  \label{eq:force}
  \Fopk 
= \frac{i}{\hbar} [\Hat{p}, \Hat{H}] 
=i\hbar \left[ \nabla g({\vect x}_k)\,
    \sop_k^\dagger \aop - \nabla g^*({\vect x}_k)\, \aop^\dagger
    \sop_k \right] \;.
\end{split}
\end{equation}
Since $\Fopk$ is normally ordered, its expectation value is easily
obtained upon insertion of the stationary solution
\eqref{eq:ss_mean_a_and_sigma} of the internal variables.  For a
moving atom this expression is only approximately valid: there will be
a time lag in the internal dynamics with respect to the atom's current
position, and hence we will include corrections to $\Fopk$ to first order
in the atomic velocities.

The slow evolution of the centers of mass of the atoms, smoothed out
on the timescale $\tau \approx \mathrm{max}\{ 1/\kappa, 1/\gamma\}$ 
is described by the coupled Langevin equations:
\begin{subequations}
  \label{eq:CM_motion}
\begin{align}
  \dot {\vect x}_k &= {\vect p}_k/M, \\
  \dot {\vect p}_k &= {\vect f}_k + \sum_{m=1}^{N} {\boldsymbol
    \beta}_{km} {\vect p}_m/M + {\boldsymbol \Xi}_k \;.
\end{align}
\end{subequations}
In these equations ${\vect f}_k=\expect{\hat{\vect F}_k}$ are the
vectors giving the steady state $v=0$ contribution of the force, while
$\bkm$ are the tensors describing the first order corrections to the
force acting on atom $k$.  ${\boldsymbol \Xi}_k$ denotes the Langevin
noise forces due to photon recoil.  They correspond to random kicks
along the cavity axis with zero average and second moments given by
$\expect{{\boldsymbol \Xi}_k {\boldsymbol \Xi}_m}={\mathcal D}_{km}$.
Note that the matrix ${\mathcal D}_{km}$, representing the strength of
the Langevin noise, depends on the time-varying atomic positions.  It
represents the quantum fluctuations of the force due to the fact
$\expect{\hat{F}_k\circ\hat{F}_m}\neq\expect{\hat{F}_k}\circ\expect{\hat{F}_m}$.
It is by the addition of the noise terms ${\boldsymbol \Xi}_k$ that we
tailor our classical force to give the same second-order expectation
values as its quantum counterpart\cite{domokos03}.

\section{Interaction channels}
\label{sec:interaction}
The Hamiltonian (\ref{eq:H}) contains no direct coupling between the
two atoms: these only arise indirectly due to coupling to the same
field mode.  Interestingly the atom-atom interaction appears in all
the three types of forces present in classical equations of motion
(\ref{eq:CM_motion}).  First, the steady state force $\vect{f}_k$
depends on the positions of both atoms via the steady state intensity.
Second, not only does the friction coefficient on one atom depend on
the position of the other, but the friction matrix has off-diagonal
terms as well. This means that apart from ordinary viscous friction
($\dot{\vect p}_k \propto \vect{v}_k$) a strange phenomenon, which we
call cross-friction ($\dot{\vect{p}}_k \propto \vect{v}_l$, for $l \ne
k$) is also present. Here the velocity of one atom influences the
friction experienced by the other atom. Third, the Langevin noise term
on one atom has an expected magnitude influenced by the position of
the other, and the noise terms ${\boldsymbol \Xi}_k$ are directly
correlated as well.  Hence we get joint ``kicks'' on both atoms
leading to correlated motion. In the following we will analyze these
interaction channels in more detail.

\subsection{Steady state force $\vect{f}$}
Formally the expectation value $\mathbf{f}_k$ of the force operator
looks very similar to the case of free-space Doppler cooling:
\begin{equation}
    \label{eq:cavity_pumping_force}
{\vect f}_k = -\hbar \frac{\Delta_A}{\Delta_A^2 +
\gamma^2}\expect{\aop^\dagger\aop} \vgrad_k g^2(\xk).
\end{equation}
However it additionally depends on the positions of the other atoms
via the cavity field intensity.  This dipole force is conservative and
can be derived from a potential.  The potential looks more complicated
than in free space as it contains the dynamical nature of the cavity
field, but still can be given in closed form \cite{fischer01} :
\begin{equation}
    \label{eq:potential}
V = \frac{\hbar \Delta_A \abs{\eta}^2}{\Delta_A \kappa + \Delta_C\gamma}
\mathrm{atan}
\frac{\gamma\kappa - \Delta_A\Delta_C + \sum_l{g^2(x_l)}}
{\Delta_A \kappa + \Delta_C\gamma}.
\end{equation}

\begin{figure}
\begin{center} 
\includegraphics[angle=270,width=8cm]{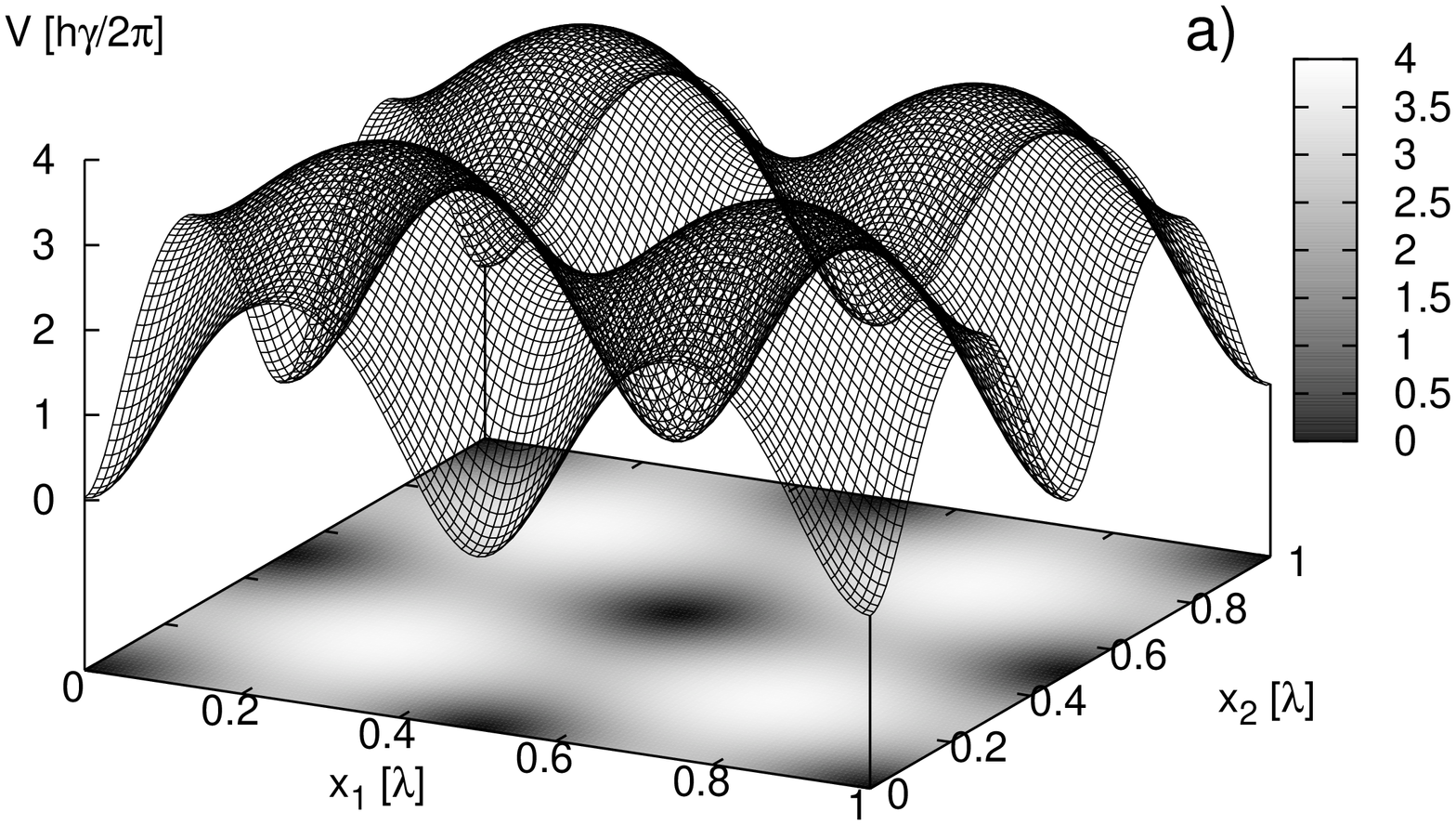}
\includegraphics[angle=270,width=8cm]{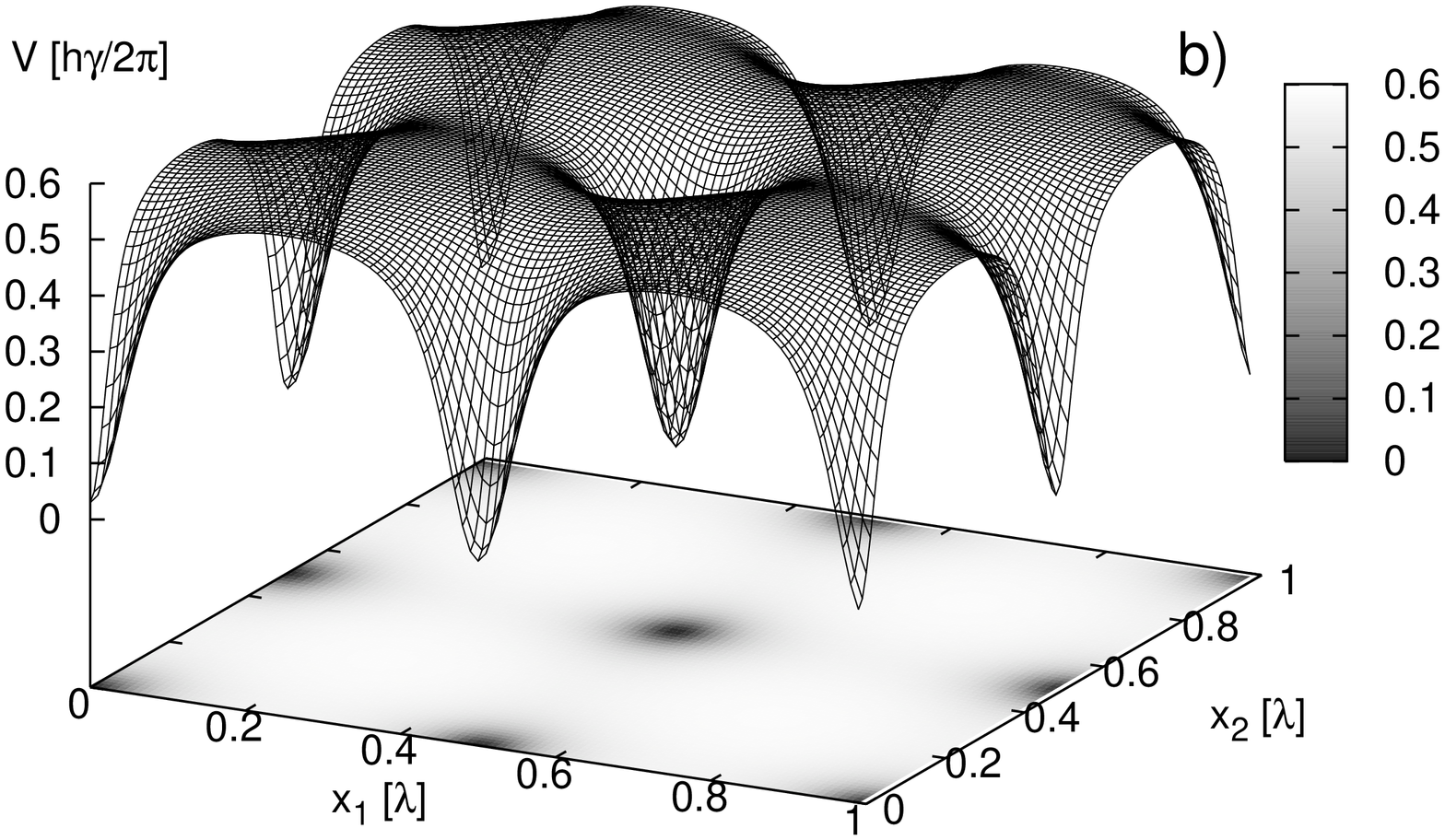}
\caption{\emph{Potential as a function of atomic positions} 
The potential (\ref{eq:potential}) is plotted as a function of
positions of the two atoms. In a) the Garching parameters 
($\kappa=\gamma/2$,$g_0=5\gamma$)
are used, in b) $g_0$ was increased fourfold. 
In the first case, the
potential is well approximated by a sum of two single-particle
potentials. In the second case, it is seen that 
either both atoms are trapped or both are free.
The trap is deeper for smaller couplings. 
\label{fig:potentials} 
}
\end{center}
\end{figure}

To show the effects of dynamic field adjustment, we plot this
potential for two typical cases in Fig.~\ref{fig:potentials}.  For the
experimental parameters used in an experiment at MPQ in Garching
\cite{fischer02} (Garching parameters), $\kappa=\gamma/2$,
$g_0=5\gamma$, and for a detuning $\Delta_A=-50\gamma$ (upper graph)
the effective interaction between the atoms is relatively weak and the
potential resembles the familiar ``egg-carton'' surface proportional
to $\sin^2(k_C x_1)+\sin^2(k_C x_2)$.  For a somewhat stronger atom-field
coupling $g_0=20\gamma$ (lower graph in Fig.~\ref{fig:potentials}) the
atomic interaction is quite obvious and the shape of the potential of
the second atom strongly depends on the position of the first atom and
vice versa. Basically in this second case either both atoms are
trapped, or both are free.

The peculiar $g_0$-dependence of the interaction and the trapping
effects can be understood physically by looking at formula
(\ref{eq:cavity_pumping_force}).  The atoms see each other through the
cavity field $\expect{\hat a^\dag \hat a}$.  As we saw in formulae
(\ref{eq:rezonancia_D},\ref{eq:U0_def},\ref{eq:Gamma0_def}), the field
is in resonance when $U({\vect x}_1,{\vect x}_2)$ is approximately
$\Delta_C \pm (\Gamma+\kappa)$.  Each atom can shift $U$ in this
far-detuned case by approximately $U_0\approx g_0^2/\Delta_A$, whereas
the cavity linewidth is approximately $\kappa + \gamma
g_0^2/\Delta_A^2$.  Using the MPQ parameters $U_0 < \kappa$, and
therefore the back-action of the atoms on the cavity field is weak.
In the large-$g_0$ case, $U_0 > \kappa$, but $\Gamma_0<\kappa$,
meaning that the atoms can shift the cavity resonance significantly
more than a linewidth. Hence if one of the atoms leaves its trap it
will shift the cavity out of resonance and cause the other atom to be
released as well. Moreover, the amplitude of the force is proportional
to the cavity field, which in the large-$g_0$ case decreases faster
with the distance from the trapping point. This implies that less work
needs to be done to free an atom: the potential is smaller, despite
the same maximum light shift.

Let us now analyze the interaction in a more quantitative way.  We
start from the limiting case of neglecting the atomic back-action on
the field (i.e.~we assume a constant field intensity). This 
corresponds to the force \eqref{eq:cavity_pumping_force}
\begin{equation}
    \label{eq:cavity_pumping_force_large_detuning}
{\vect f}_l \approx \frac{\hbar \abs{\eta}^2}{2\Delta_A \kappa^2}
\vgrad_l g^2(\xl)
\end{equation}
one obtains in the limit of very large atom-field detuning. 
We can now find corrections from the cavity-mediated interaction 
to this force by expanding the potential \eqref{eq:potential} in 
a power series. 
Setting the driving field to resonance,
$\Delta_C=U_0-\kappa$, the potential Eq.~\eqref{eq:potential} to
second order in ${\gamma}/{\Delta_A}$ reads
\begin{multline}
    \label{eq:potential_power_series}
V \approx \frac{\hbar \abs{\eta}^2}{\kappa} \left[
\frac{\pi}{4} - \left( \left( 1 + \frac{\pi}{4} \right) 
+ \frac{g_0^2}{2\kappa\gamma}
\sum_{l=1}^N {f^2({x}_l)} \right) \frac{\gamma}{\Delta_A} \right.\\
\left.+ \Bigg\{ \left( 1 + \frac{\pi}{4} \right)
    - \left( 1 + \frac{\pi}{2} \right) 
N \right. \\
\left.+ \left(\frac{g_0^2}  {2\kappa\gamma}\right)^2
\bigg( \sum_{l=1}^{N}{f^2(x_l)} \bigg)^2 \Bigg\} 
\left(\frac{\gamma}{\Delta_A}\right)^2
\right] .
\end{multline}
To first order in the small parameter we have a sum of single-atom
potentials, giving the ``egg-carton'' shape. The corrections to this
are given by terms of higher order in our expansion parameter, of
which we give the first nontrivial term here.  Note that since it is
not simply the distance of the atoms upon which the potential depends,
the interatomic force between them -- as can be read out from the
above formula -- is not a ``force'' in the sense of Newton's third
law.

\subsection{Forces linear in velocity: friction and cross-friction}
To lowest order in the adiabatic separation of the internal and
external dynamics we used the steady state values of the internal
variables for fixed positions of the atoms to calculate the above
potentials.  As a next step we can include corrections for $\hat a$
and the $\sop_k$ linear in the velocity ${\vect v}_m$ of each atom,
which should be valid for low velocities.  As described in
\cite{domokos03} this leads to a friction matrix $\bkm$ as first order
correction to ${\vect f}_k$.

We obtain the following explicit formula for the friction matrix,
\begin{multline}
   \label{eq:beta_simpler}
   {\boldsymbol \beta}_{km} = 
   2 \hbar \vgrad_k g(\xk) \circ 
   \vgrad_m g(\vect{x}_m) \frac{\eta^2}{\abs{\dd'}^2} \gamma
   \frac{2\Delta_A}{\Delta_A^2+\gamma^2} \delta_{km}\\ 
   + \hbar \vgrad_k g^2(\xk) \circ 
   \vgrad_m g^2(\vect{x}_m) \frac{\eta^2}{2\abs{\dd'}^2} 
   \Im \left\{ \frac{1}{{\dd'}^2} \cdot \right. \\
   \left. \left( 2(1 + \chi) \left((i\Delta_A-\gamma)^2 - \sum_l g(\xl)^2 \right)   
       + (1 + 3\chi) {\dd'}\right) \right\} 
\end{multline}
where $\chi=(i\Delta_A+\gamma)/(i\Delta_A-\gamma)$ is a complex factor
of unit modulus, which for large atomic detuning, $\Delta_A\gg\gamma$,
becomes approximately $\chi\approx 1$.  This formula differs somewhat
from that obtained by Fischer et al.~\cite{fischer01} by a slightly
different approach. The most important difference is that we find a
matrix \emph{symmetric} in the indices ${km}$. This is important because it is these 
off-diagonal terms that couple the velocities of the atoms, and 
have a decisive influence on the buildup of correlated
motion. We defer detailed discussion of the formula 
\eqref{eq:beta_simpler} to Section \ref{sec:analytical}.

\subsection{Random forces due to quantum noise} 
Spontaneous emission and cavity decay introduce quantum noise into the
atomic motion.  These heat up the system and generally tend to
decrease motional correlations of the atoms.  Following the line of
reasoning briefly mentioned at the end of section \ref{sect:themodel}
and discussed in more detail in \cite{domokos03}, we can calculate the
influence of the noise operators $\hat \xi$ and $\hat \zeta$ of
eq.~(\ref{eq:bosonic_Langevin}) on the dynamics.  For $N$ atoms we
arrive at the following simple formula:
\begin{multline}
    \label{eq:d_simpler}
{\dd}_{km} = 2 \hbar^2 \vgrad_k g(\vect{x}_k) \circ 
\vgrad_m g(\vect{x}_m) \frac{\eta^2}{\abs{\dd'}^2} \gamma
\delta_{km} \\
+ 2 \hbar^2 \vgrad_k g(\vect{x}_k)^2 \circ \vgrad_m g(\vect{x}_m)^2 
\frac{\eta^2}{\abs{\dd'}^2} 
\Delta_A 
\frac{\kappa \Delta_A + \gamma \Delta_C }{\abs{\dd'}^2}.
\end{multline}
This is a simple extension of the corresponding formula for one atom
given in \cite{domokos03}. Let us remark here, that the diagonal part
of this diffusion matrix ${\dd}_{km}$ has been also found by Fischer
et al.~\cite{fischer01}.  Surprisingly one also obtains off-diagonal
terms, which have not been considered before.  These terms lead to
correlated kicks on the atoms, which can .add to the atom-atom
correlations, rather than destroying them.

Spontaneous emission adds recoil noise, which gives an extra term 
to the noise correlation matrix of the form:
\begin{equation}
    \label{eq:d_spontan}
{\dd}^{\mathrm{sp}}_{km} = \delta_{km} 2 \hbar^2 k_A^2 \overline{u^2} 
\frac{\eta^2 g(\vect{x})^2}{\abs{\dd'}^2} \gamma.
\end{equation}
Here $k_A=\omega_A/c$ is the expected wavenumber of the emitted
photons and  $\overline{u^2}$ is the correction factor coming from the
spatial distribution of the photons, in our case $\overline{u^2}=2/5$.
As expected, spontaneous emission, being a single-atom process, induces no
correlations between the atoms.

\section{Numerical simulations of the correlated atomic motion}
\label{sec:numerical}
Having discussed the qualitative nature of the combined atom field
dynamics, we now turn to numerical simulations for some quantitative
answers.  We numerically integrate the Langevin equations
(\ref{eq:CM_motion}), varying the ratio of the cavity loss rate to the
linewidth of the atom $\kappa/\gamma $ as well as the relative
coupling strength $g_0/\gamma$. Note that the relative magnitude of
radiation pressure and dipole force can be changed by varying the
detuning between pump frequency and the atomic resonance.  As we are
interested in cooling and trapping the atoms we fix the cavity
frequency at $\Delta_C=N U_0-\kappa$ to ensure efficient cooling
\cite{horak97,domokos03}.  The pump power is always chosen to keep the
atomic saturation low and approximately at ($\expect{\sop_k^+
  \sop_k}<0.1$) at all times.

\begin{figure}
\begin{center} 
\includegraphics[width=4.2cm]{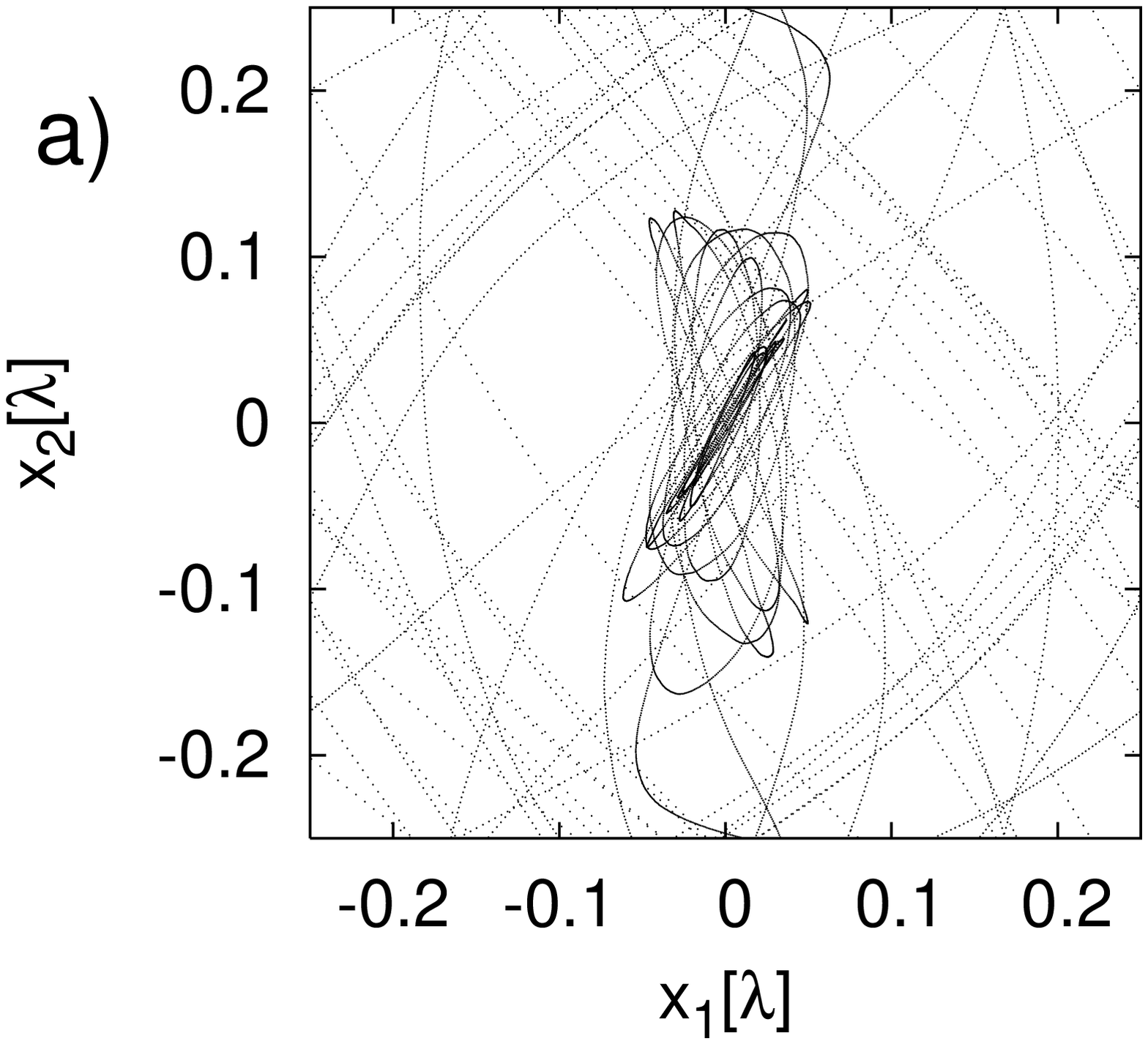}
\includegraphics[width=4.2cm]{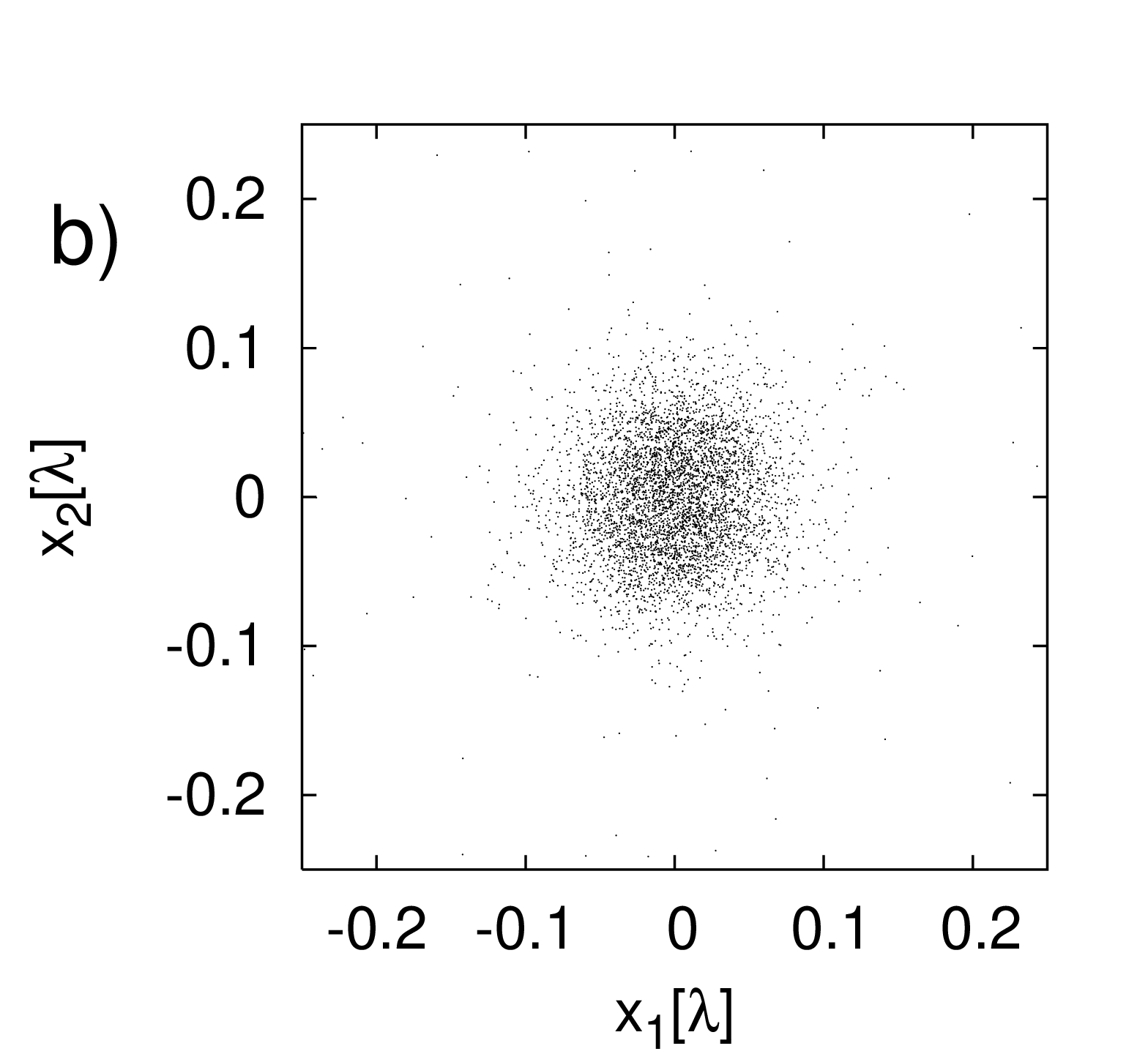}
\includegraphics[width=4.2cm]{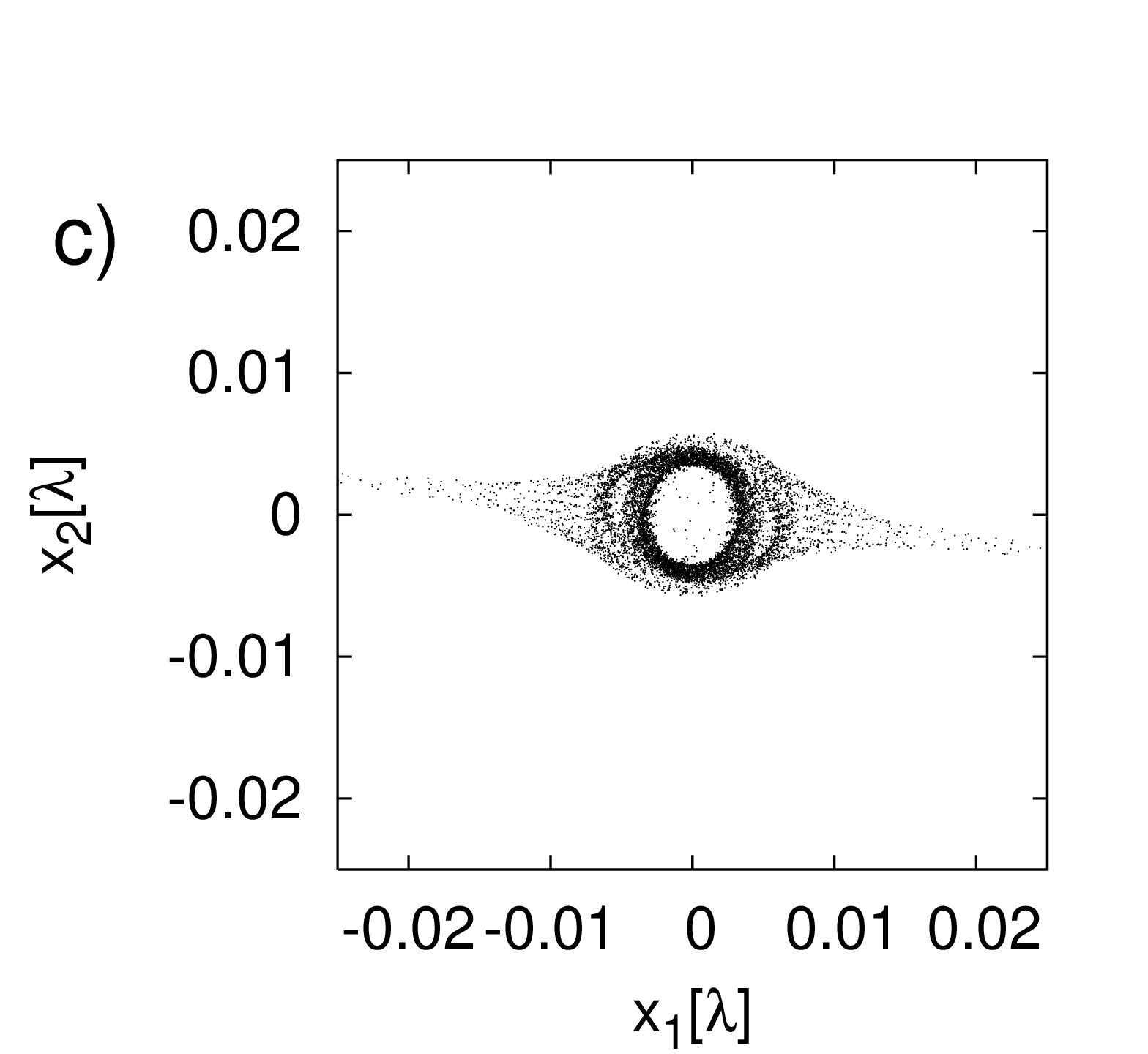}
\includegraphics[width=4.2cm]{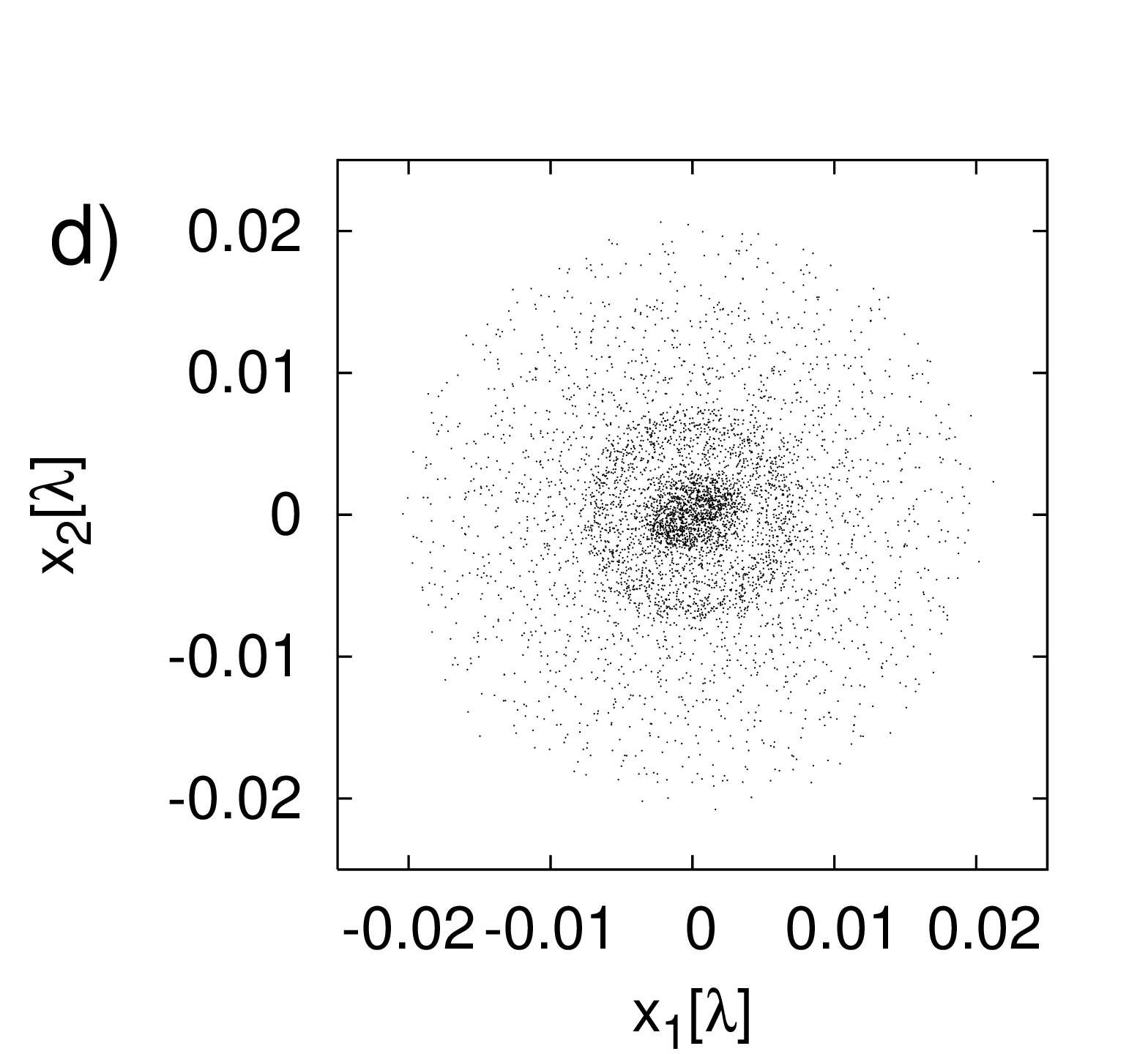}
\caption{\emph{Typical trajectories in coordinate space.} 
In the left column, in a) and b), the parameters of the Garching group are used, 
in the right column, in c) and d), we considered a better resonator 
($\kappa=0.1\gamma, g_0=10\gamma$) 
and larger detuning ($\Delta_A=-10000\gamma$).
In both cases the first $50\mu \mathrm{s}$, in a) and c), and the first 
$3\mathrm{ms}$, in b) and d), are shown.
The coordinate is the distance from the nearest trappping point. 
\label{fig:circles} 
}
\end{center}
\end{figure}

Typical trajectories of atom pairs are shown in 
Figure \ref{fig:circles}.
The atomic positions $\big( x_1(t_n), x_2(t_n) \big)$ are plotted at
regular time intervals $t_n$.  For better visibility the coordinate is
always measured from the nearest trapping point.  In the left column
the detuning was chosen ($\Delta_A=-50\gamma$), with the cavity
parameters of the MPQ group at Garching \cite{fischer02}
($\kappa=\gamma/2$, $g_0=5\gamma$).  The first $50\mu \mathrm{s}$ are
displayed in the figure on the top, and the first $3\mathrm{ms}$ in
the one on the bottom.  Both atoms localize during the first $50\mu
\mathrm{s}$ to within 1/4 of a wavelength around respective trapping
centers, a sign of cooling and trapping by the cavity.  In the right
column, the detuning is chosen much larger ($\Delta_A=-10^4\gamma$)
with stronger atom-field coupling ($\kappa=\gamma/10$,
$g_0=10\gamma$).  In this case the relative importance of spontaneous
emission is strongly reduced.  The cooling in the cavity is faster and
both atoms reach a steady state rapidly.

The appearance of a circular structure is the striking feature of the
right column.  This indicates that the motion of the atoms is
correlated.  Both atoms move sinusoidally about their respective
trapping points, with some noise but in such a way that the relative
phase of the two oscillations is likely to be $+90^\circ$ or
$-90^\circ$.  

To quantify the correlation between the atomic oscillators we define
their oscillator phases.  This is computed in the simulation using the
trap frequency, which is:
\begin{equation}
\omega_{\mathrm{trap}} = \sqrt{2 \hbar \left|\Delta_A\right| 
\expect{\sop^\dag \sop} k_C^2 / M},
\end{equation}
if both atoms are well trapped.  Here $k_C$ is the resonator mode
wavenumber, and $\expect{\sop^\dag \sop}$ is the
saturation of either atom at the trapping point.  This formula can be
derived by expanding the potential (\ref{eq:potential}), and
substituting our particular choice of $\eta$ and $\Delta_C$.

\begin{figure}
\begin{center}
\includegraphics[width=4.2cm]{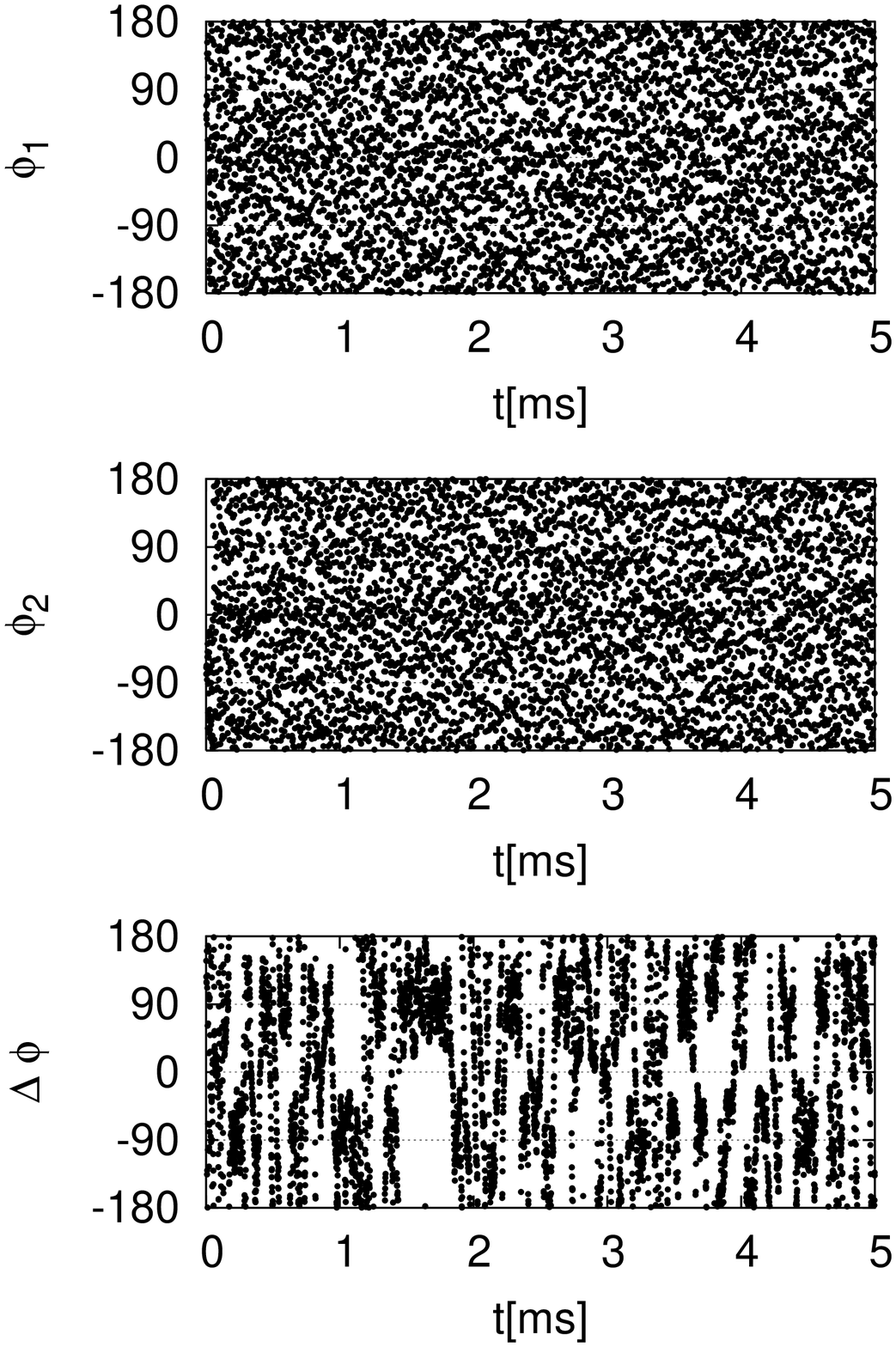}
\includegraphics[width=4.2cm]{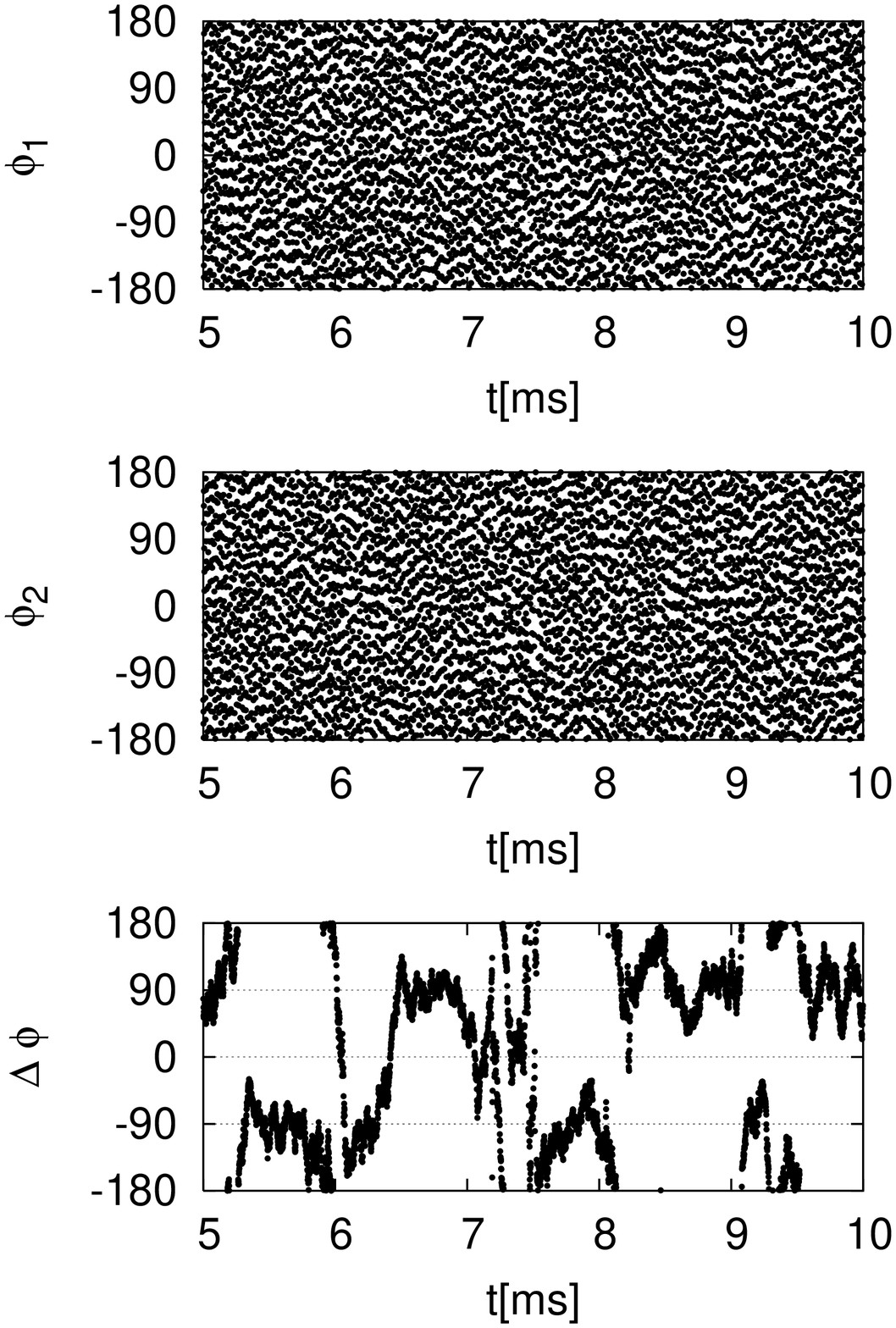}
\caption{\emph{The phases of the atomic oscillations and the relative
    phase.}  The time evolution of the oscillator phases of the first
  (upper row) and second (middle row) atoms is shown, with the phase
  difference (bottom row).  On the left the parameter set of the
  Garching experiments is used, on the right a better cavity is taken
  with larger atomic detuning, as in the text.  The phases of the
  atoms evolve too fast on this timescale, only noise is seen.  The
  phase difference is slower, with the ''improved'' parameter set it
  appears to stabilize at $+90^\circ$ and $-90^\circ$.
\label{fig:time_phases} 
}
\end{center}
\end{figure}

The measured time evolution of the 
oscillator phases for the MPQ parameters (left column)
and the ``improved'' parameters (right column) are shown in 
Fig.~\ref{fig:time_phases}. 
For each parameter set, representative runs shown in
Fig.~\ref{fig:circles} are used, and the oscillator phases of atom 1
(upper row), atom 2 (middle row) and the phase difference (lower row)
are displayed.  The rapid oscillation of the atoms (the period is
$2.9\ \mu$s for the Garching parameters and $0.17\ \mu$s for the
idealized ones) means that the time evolution of the phases is too
fast to be followed on the timescale shown, in the upper and middle
rows only ``noise'' is seen.  The phase difference, however, evolves
more slowly. This effect is more pronounced for the second parameter set,
where the dipole force dominates.  Moreover, in this second case, the
phase difference clearly stabilizes around $+90^\circ$ or $-90^\circ$,
with random jumps in between, as expected from the results shown in
Fig.~\ref{fig:circles}.

\begin{figure}
\begin{center}
\includegraphics[width=8.5cm]{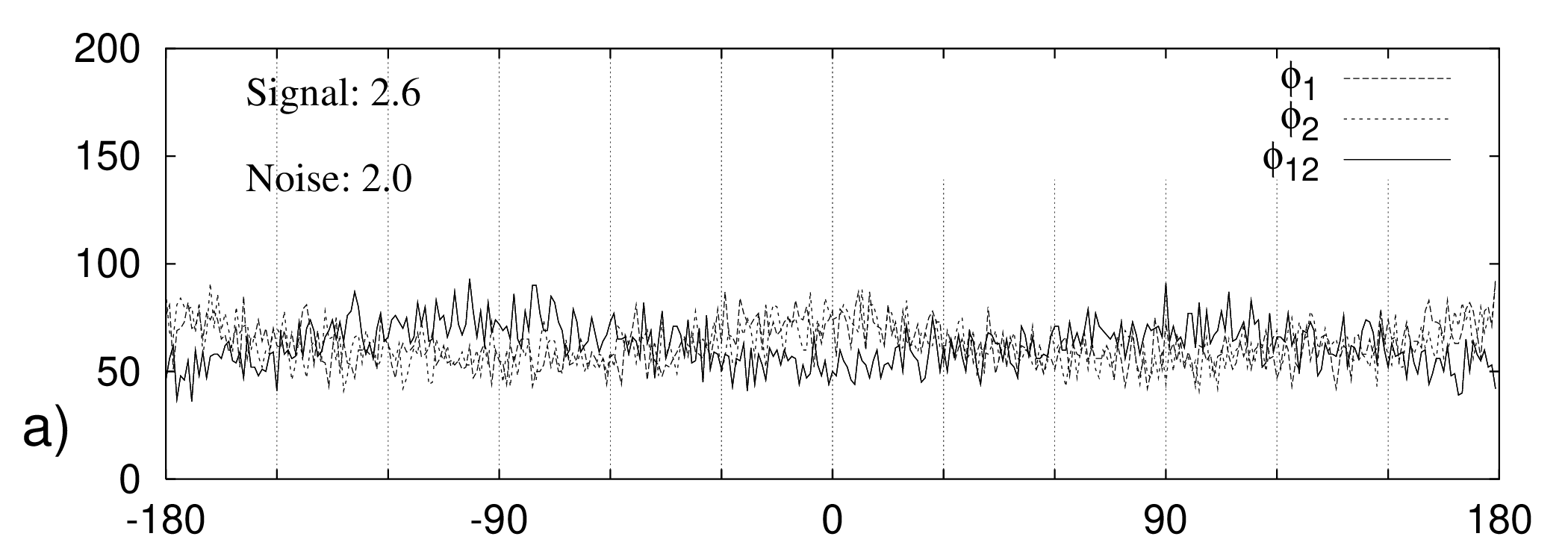}
\includegraphics[width=8.5cm]{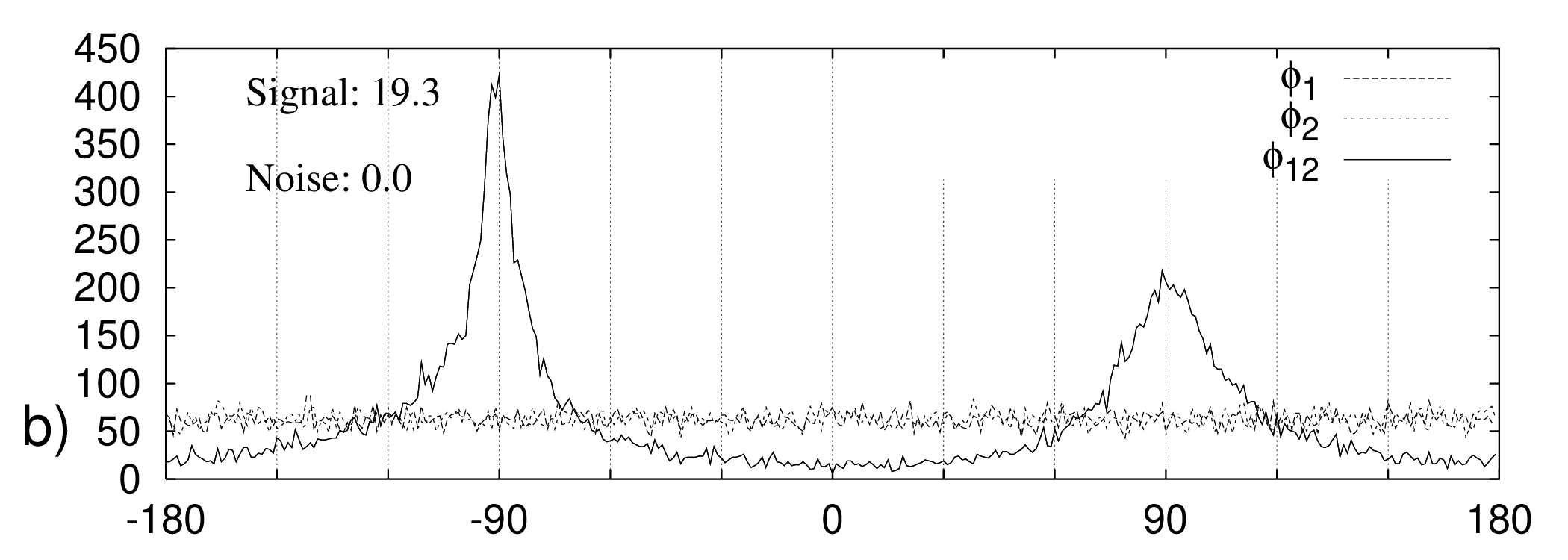}
\caption{\emph{The distribution of the phases ($\Phi_1$, $\Phi_2$) 
and the relative phase ($\Phi_{12}$).}
  For the Garching parameter set, in a), the distributions are
  flat.  Slight peaks in $\Phi_1$ and $\Phi_2$ at
  $0^\circ$ and $\pm180^\circ$ signal fast and free atoms. 
  With the improved parameters, in b), the trapping is better.
  Strong peaks at $\pm90^\circ$ in $\Phi_{12}$ indicate
  correlation of atomic motion. The strength of these peaks is
  quantified by their width, both for the phase (Noise) and the phase
  difference (Signal).}
\label{fig:distribution_phases} 
\end{center}
\end{figure}

What is left is to define a single number that quantifies the strength
of the time averaged correlations. For this we sample the
distributions of the oscillation phases and the phase difference over
time to create the histograms shown in
Fig.~\ref{fig:distribution_phases}.  As we expect, the distribution of
the phases is relatively flat for both parameters sets. However, there
is a significant difference in the distribution of the relative phase:
we find very pronounced peaks around $+90^\circ$ and $-90^\circ$ for
the second parameter set.
 
The asymmetry in these peaks is a numerical artefact due to the finite
sampling time.  It is strongly diminished if we average over several
different initial conditions.  A parameter that measures the magnitude
of the correlation is the width $W$ of these peaks, defined through
$W^2={\overline{(|\Delta\phi|-90^\circ)^2}}$, where the overbar
denotes averaging over time and over several trajectories with
different initial conditions.  Subtracting this width from the width
of the flat distribution we get the signal strength $S=51.96^\circ-W$,
which can be obtained directly during the simulation.

Let us point out here, that this measure $S$ of motional correlation
is useful only if the atoms are well trapped. In fact, fast atoms
freely moving along the lattice generate peaks in the single atom
phase distributions of $\phi_1$ or $\phi_2$ around $0^\circ$ or
$180^\circ$.  We therefore monitor the single atom distributions
simultaneously and measure the widths of the peaks around $0^\circ$
and $180^\circ$ in the same way as we did for the signal $S$.  If the
``Noise strength'' is too large, ($>10^\circ$, an arbitrary value), we
declare our correlation measure unusable.

\subsection{Scanning the parameter space}

\begin{figure}
\begin{center}
\includegraphics[width=4cm]{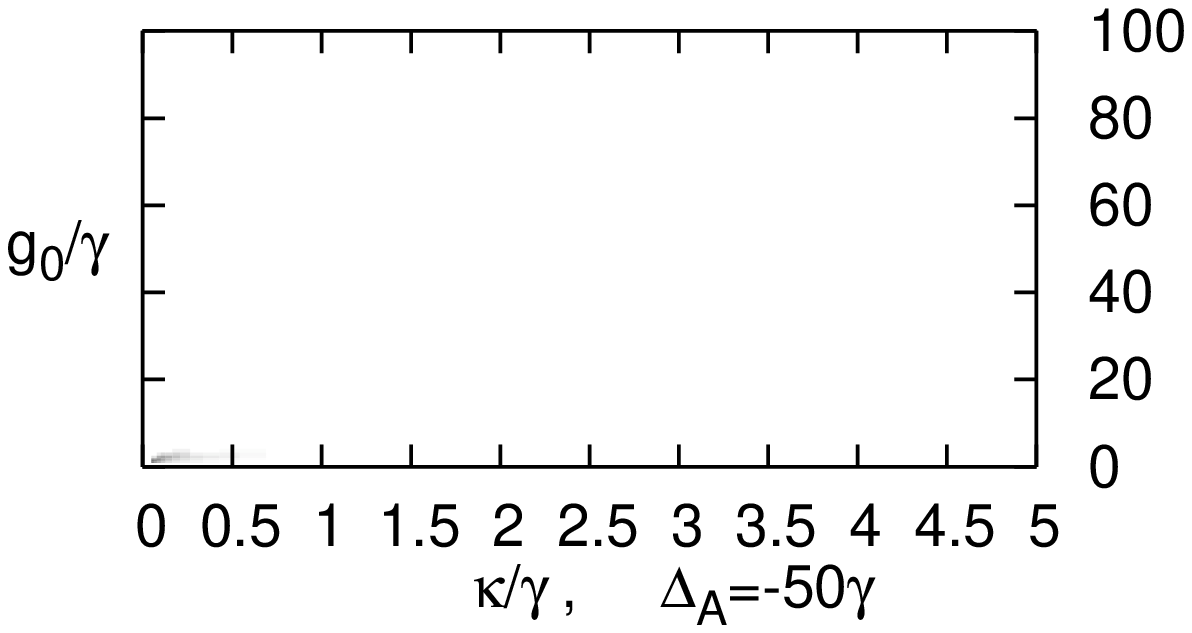}
\includegraphics[width=4cm]{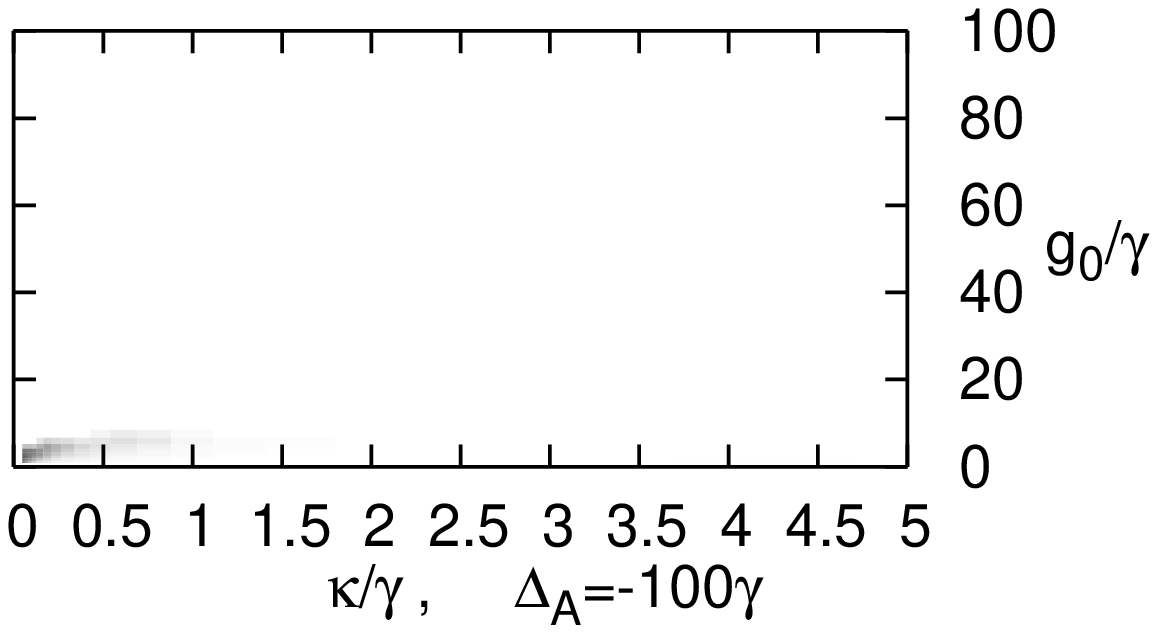}
\includegraphics[width=4cm]{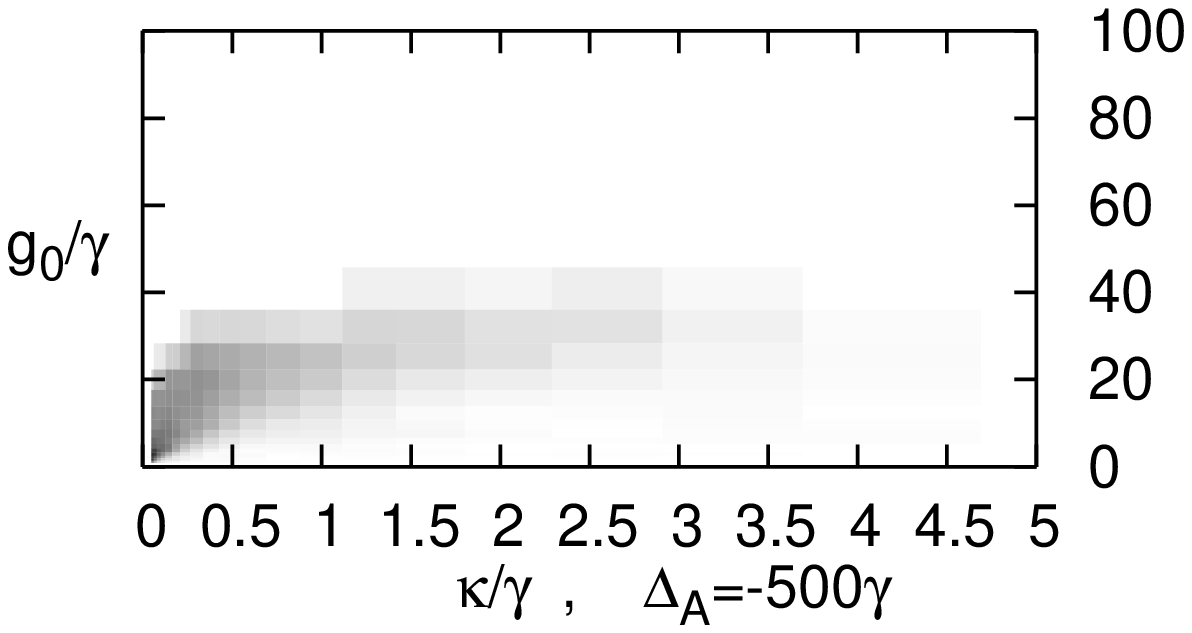}
\includegraphics[width=4cm]{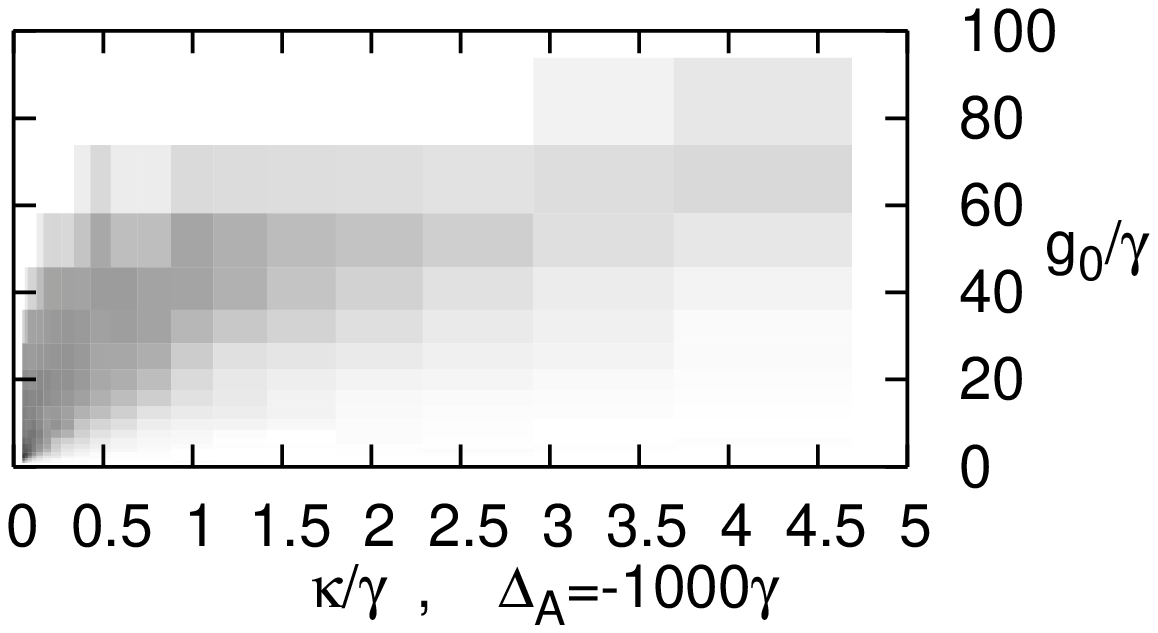}
\includegraphics[width=4cm]{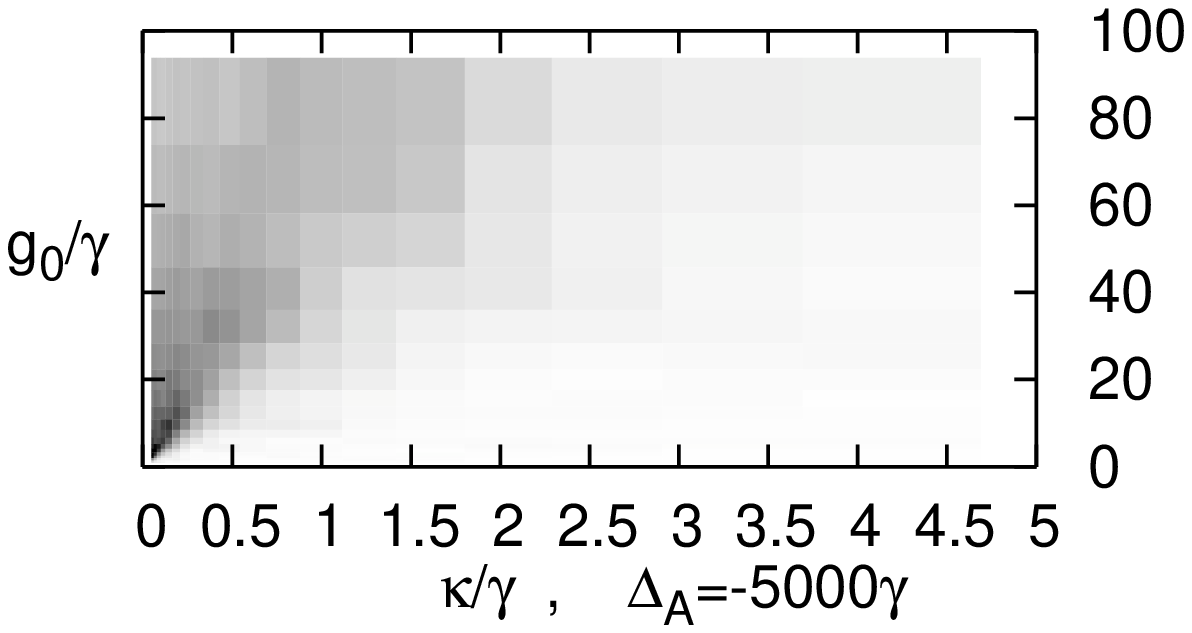}
\includegraphics[width=4cm]{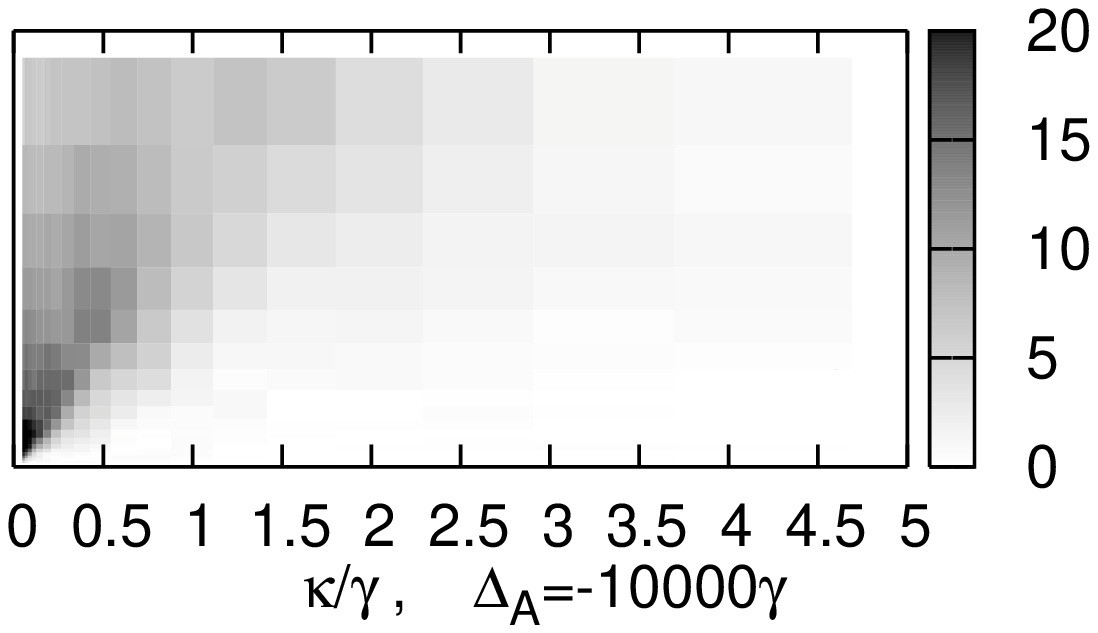}
\caption{\emph{Correlation vs $\kappa$ and $g_0$ at 
    various detunings.}  At small detunings $|\Delta_A|<100\gamma$ the
  atoms are cooled but not trapped, correlation cannot be measured.
  At higher detunings trapping is good enough, and the correlation
  strength measured as described in the article is shown in with
  shades of gray.
\label{fig:corr} 
}
\end{center}
\end{figure}

Having defined a suitable measure of correlation between the motion of
two atoms in the same cavity, we are in the position to quantitatively
investigate the parameter dependence of this phenomenon.  To this end,
we ran the simulation program for cavity decay rates of $1/50\,\gamma
< \kappa < 5\gamma\ $ and coupling strengths $\gamma < g_0 <
100\gamma$, at atomic detunings of $-10^4 \gamma < \Delta_A
<-50\gamma$ (at detunings of higher magnitude the atoms move too
rapidly and adiabaticity does not hold).  The results are plotted in
Fig.~\ref{fig:corr}.

At atomic detunings of $100 \gamma$ or less, the atoms are cooled but
generally not trapped  by the cavity, except for a small region of 
$g_0<5 \gamma$ and $\kappa<\gamma$. 
In that case the motion of two atoms will not be correlated.
For detunings as large as $5000\gamma$, the cavity field traps and
cools the atoms for any considered values of the parameters.
The correlation becomes apparent for $g_0>40\kappa$, and grows weaker
if $g_0$ is further increased. 

\subsection{Correlation and cooling}

A decisive advantage of cavity cooling of an atom is the fact that it
needs little spontaneous emission \cite{horak97}.  It has been argued
that this is a single atom effect and correlations established between
the atoms' motion will decrease efficiency or even turn off
cooling \cite{gangl99} for a trapped thermal ensemble. On the other
hand, in some simulations for very weakly bound atoms, this effect
seemed not to play any role \cite{horak01}.

\begin{figure}
\begin{center}
\includegraphics[width=4cm]{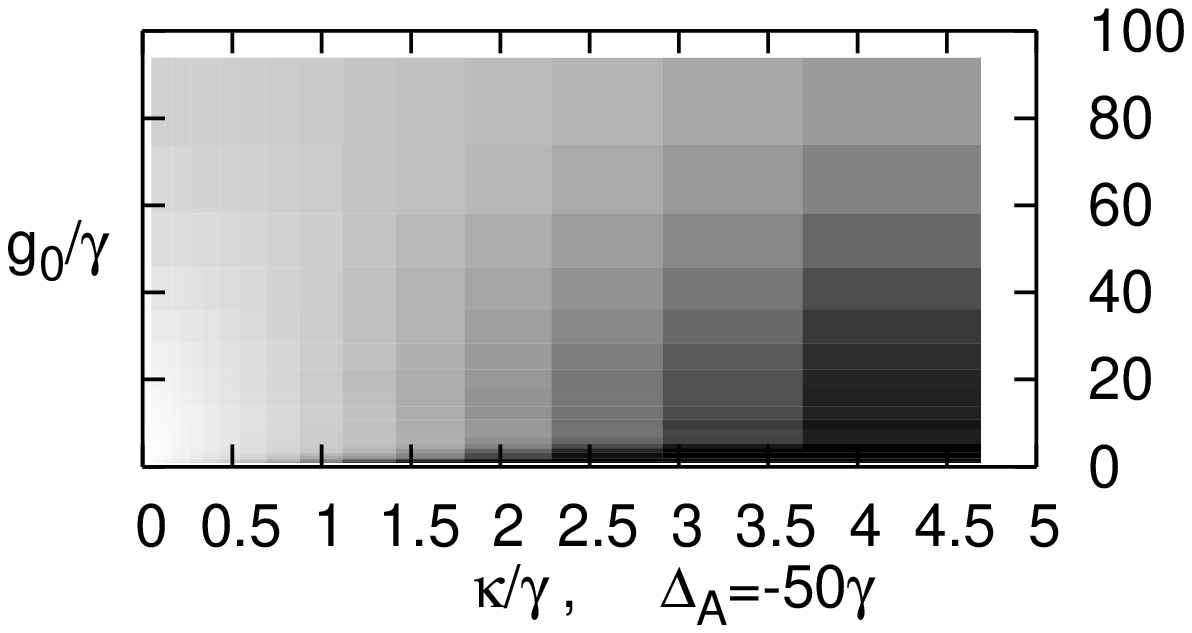}
\includegraphics[width=4cm]{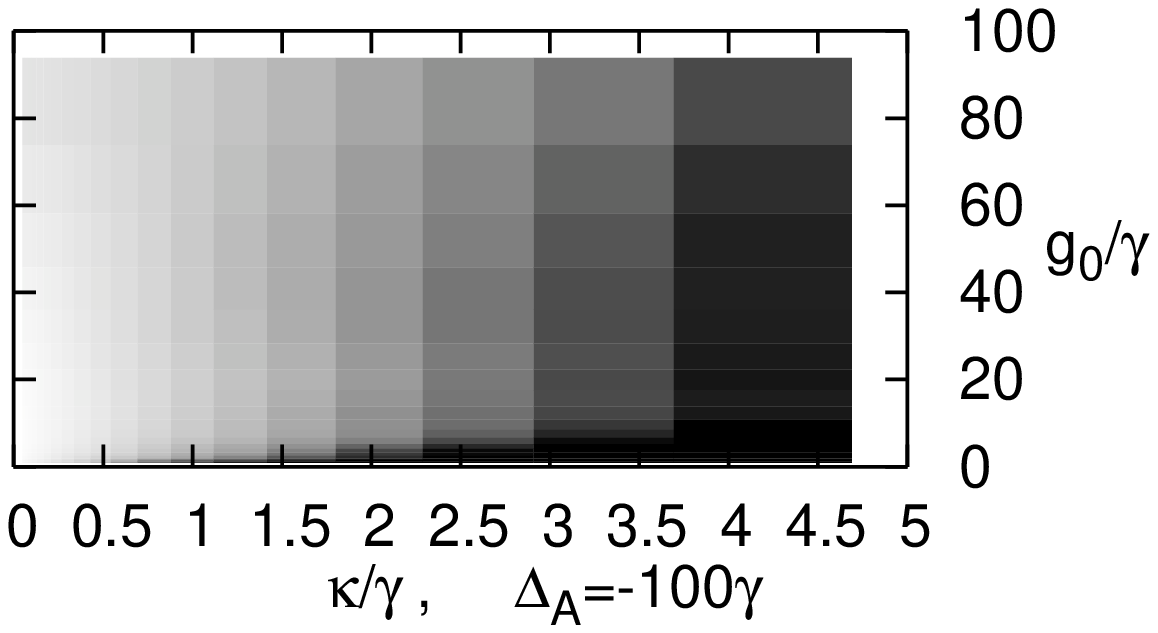}
\includegraphics[width=4cm]{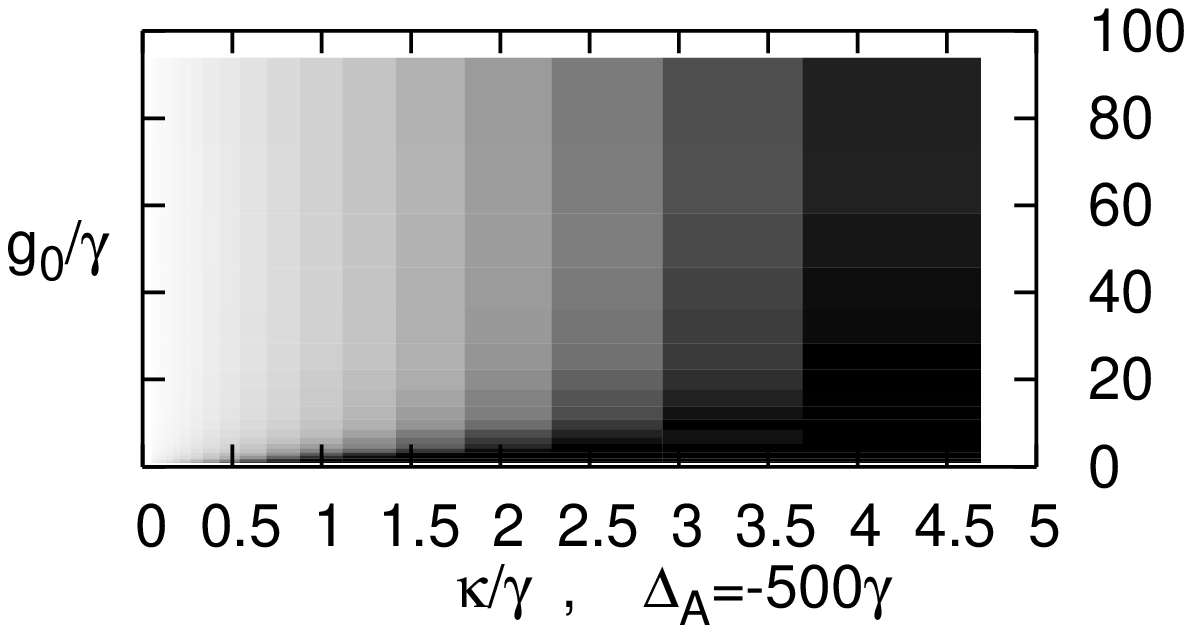}
\includegraphics[width=4cm]{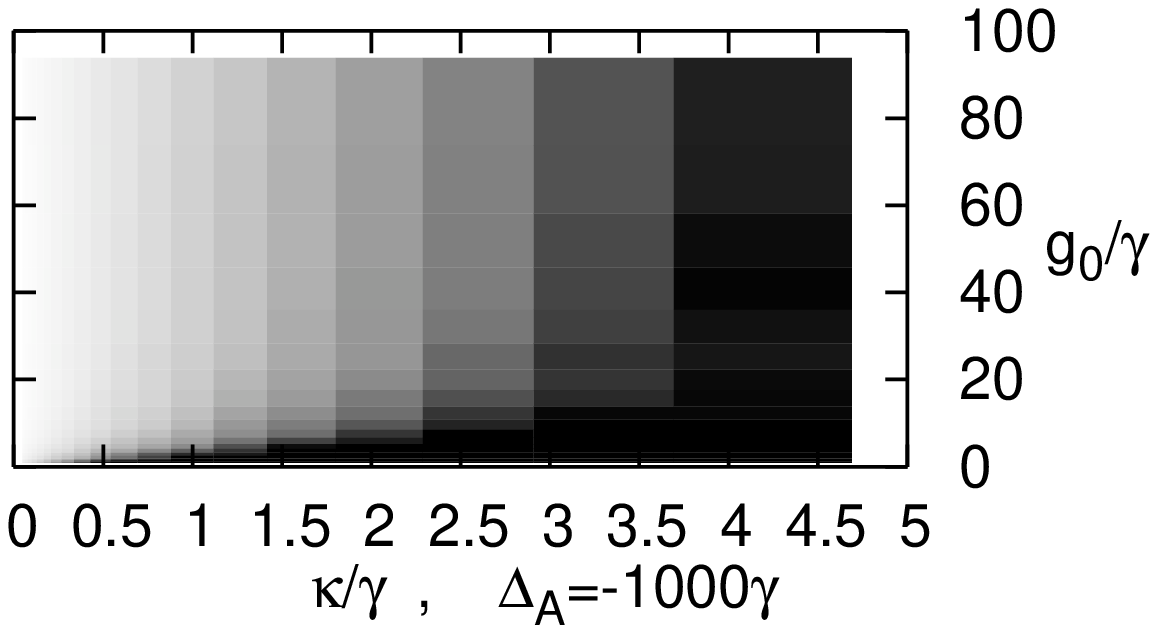}
\includegraphics[width=4cm]{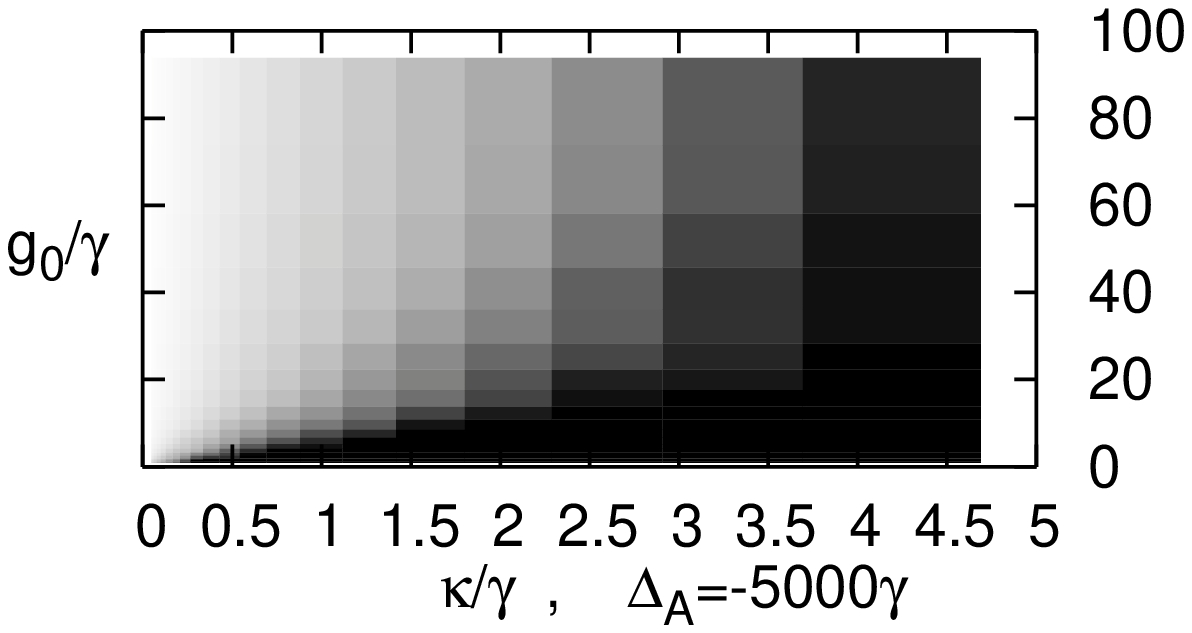}
\includegraphics[width=4cm]{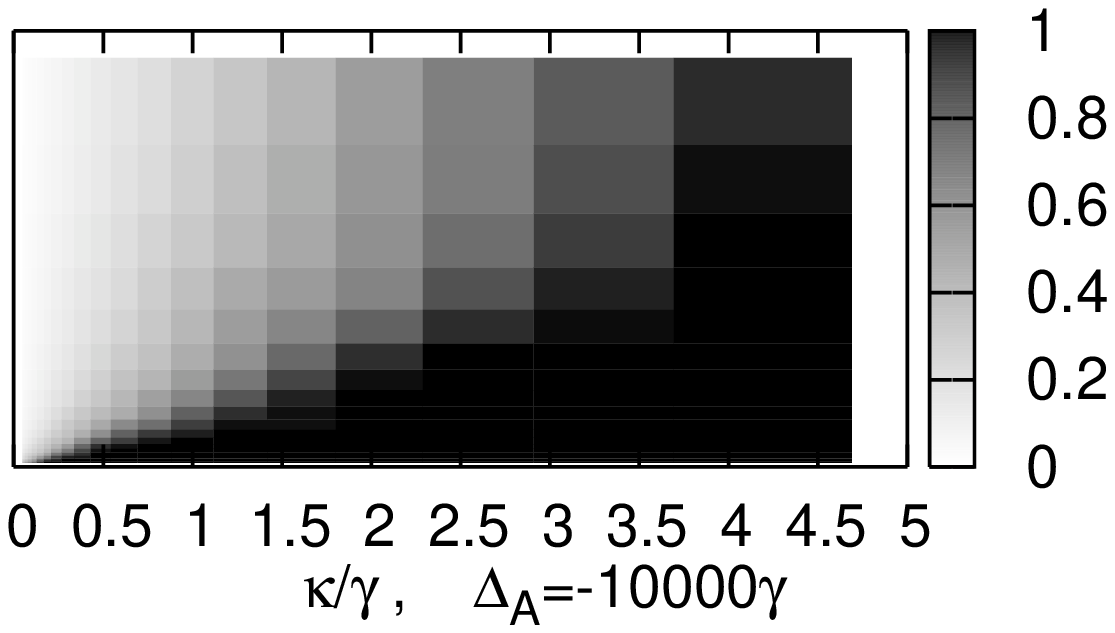}
\caption{\emph{One atom in the cavity: final temperature 
    vs.~$\kappa$ and $g_0$, for different atomic detunings.}
  Temperature is shown in units of $\hbar\gamma$ with shades of gray.
  Black denotes temperature above the Doppler limit of $\hbar \gamma$.
\label{fig:temp1} 
}
\end{center}
\end{figure}

\begin{figure}
\begin{center}
\includegraphics[width=4cm]{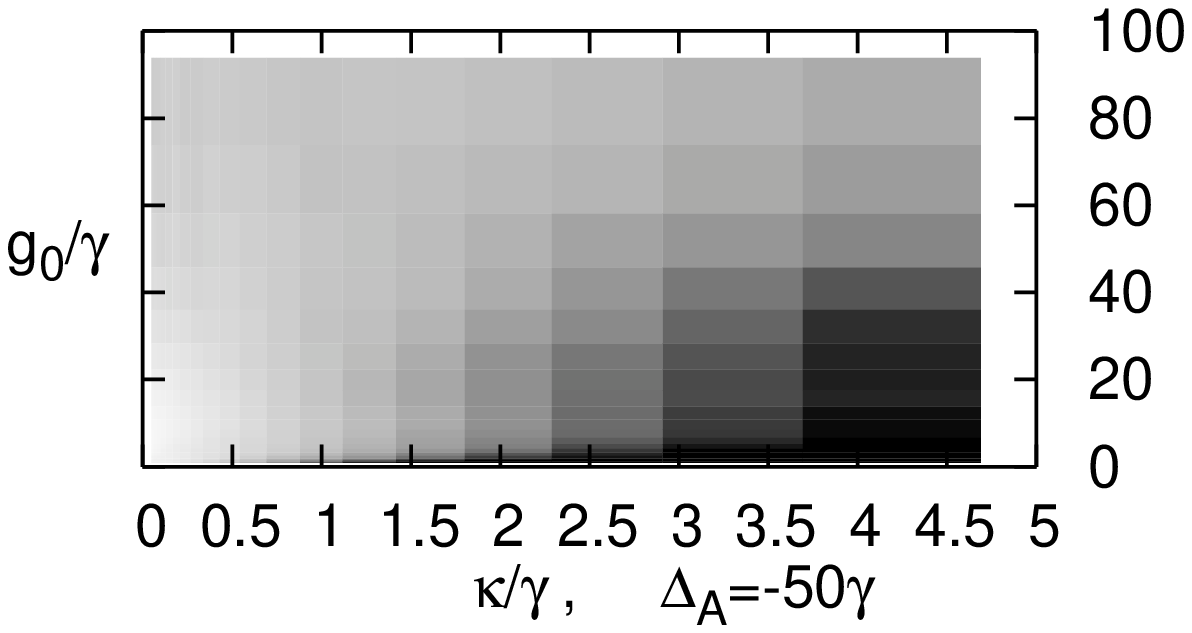}
\includegraphics[width=4cm]{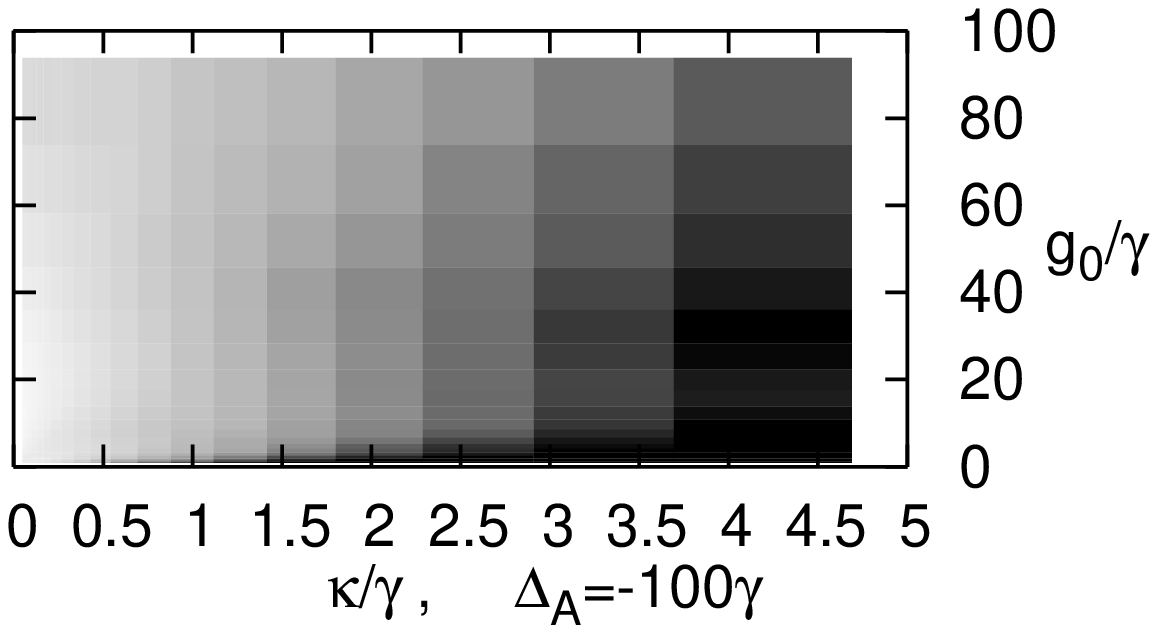}
\includegraphics[width=4cm]{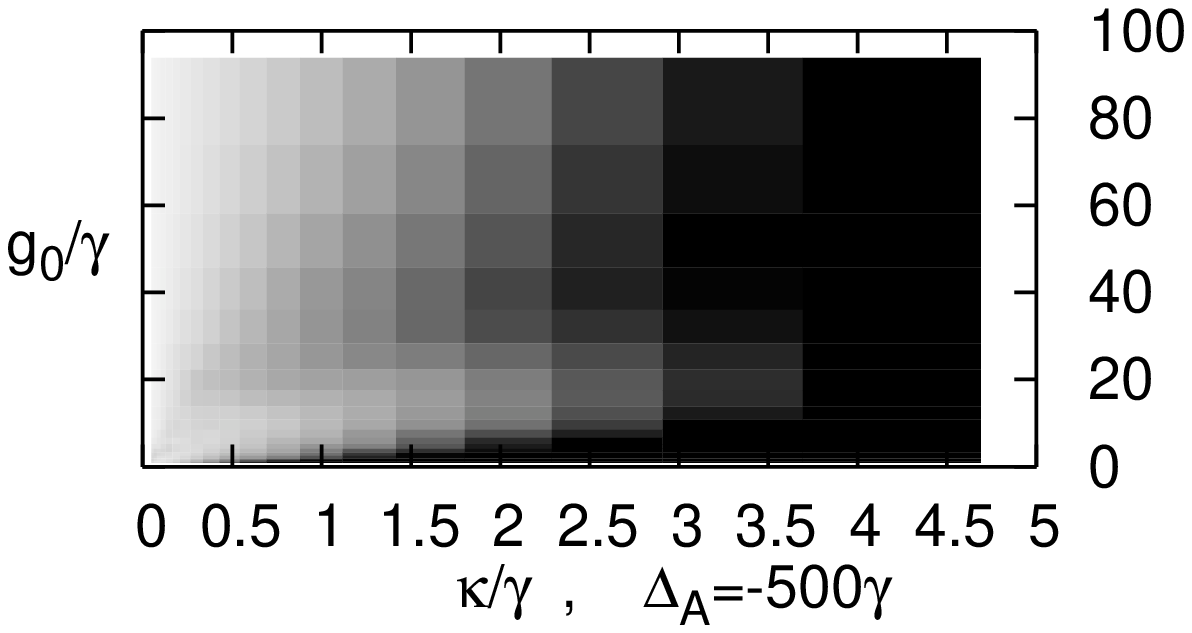}
\includegraphics[width=4cm]{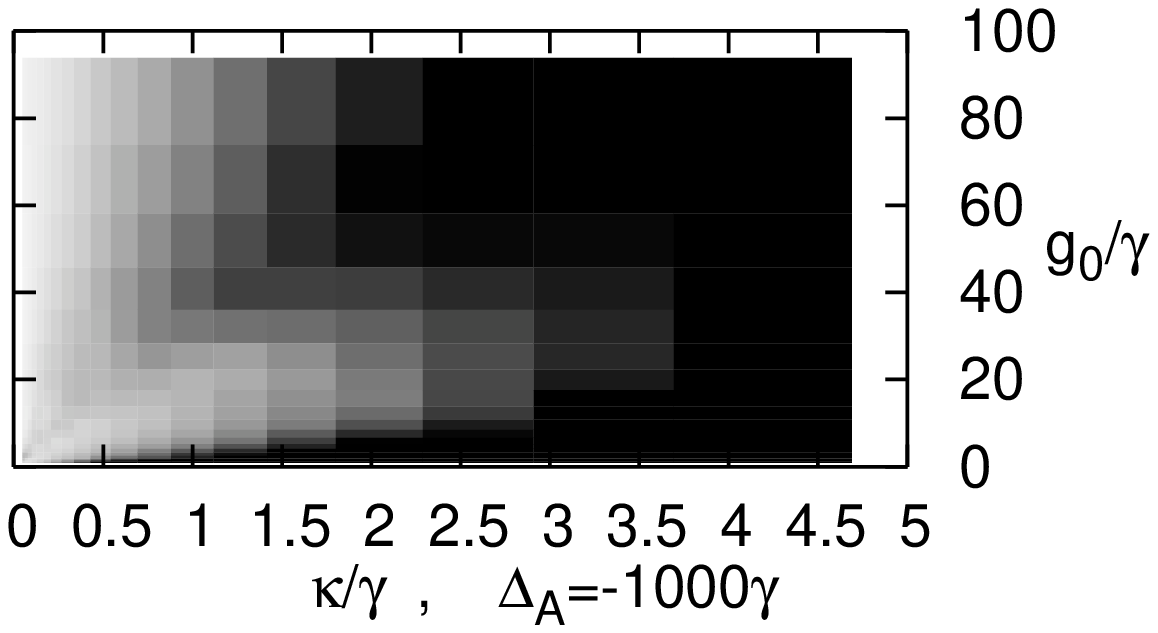}
\includegraphics[width=4cm]{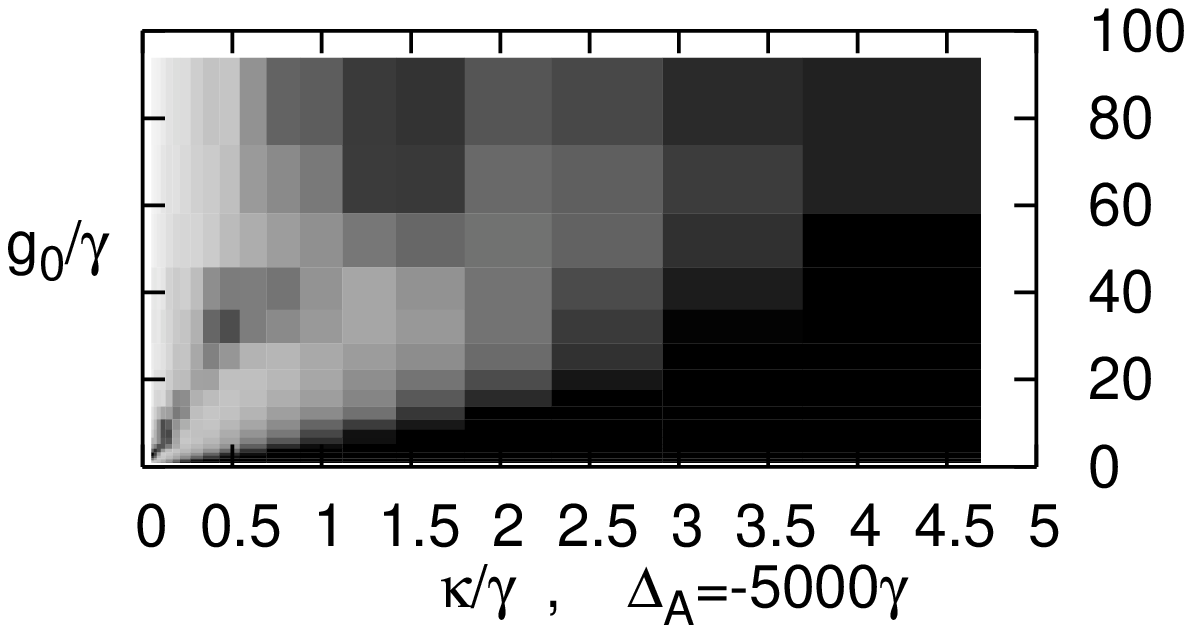}
\includegraphics[width=4cm]{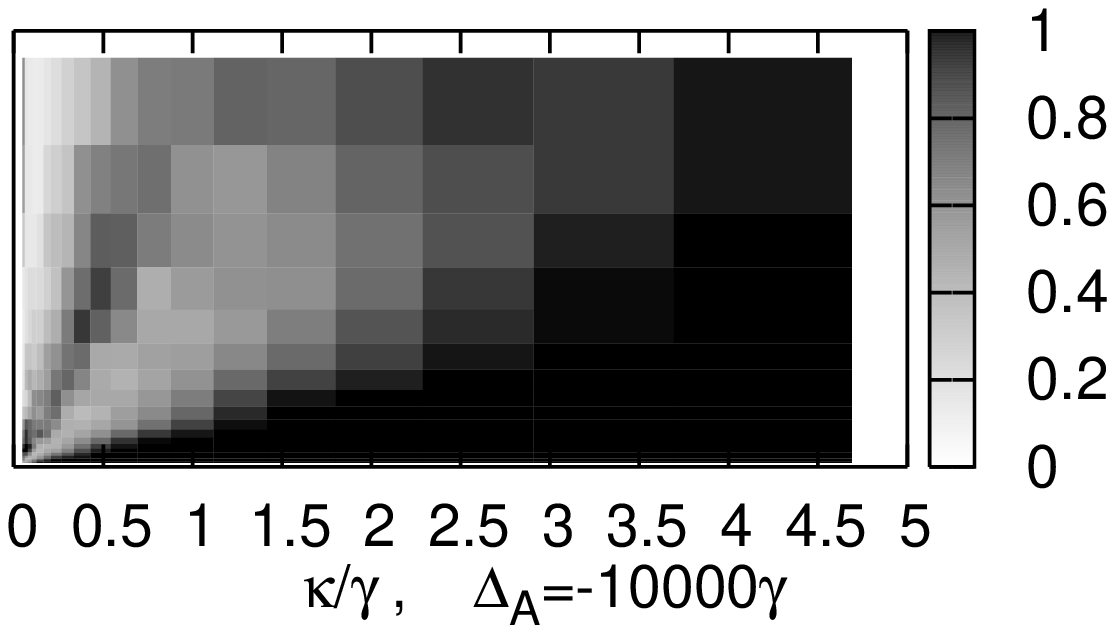}
\caption{\emph{Two atoms in the cavity: final temperature 
    vs.~$\kappa$ and $g_0$, for different atomic detunings.}
  Temperature is shown in units of $\hbar\gamma$ with shades of gray,
  as in Figure \ref{fig:temp1}.  For low atomic detuning
  $|\Delta_A|<100\gamma$ the temperature is the same as in the
  one-atom case.  For large atomic detuning new structures appear in
  the plots.
\label{fig:temp2} 
}
\end{center}
\end{figure}

In our model we can now study the effect of correlations on cavity
cooling in a very controlled way for a large range of parameters.
We performed several runs of our simulation with different cavity
parameters and at different detunings, comparing the equilibrium
temperature for one and two atoms in the cavity.  Our results are
plotted in Figs.~\ref{fig:temp1} (one atom) and \ref{fig:temp2} (two
atoms).  All shades of gray show temperatures below Doppler
temperature $\hbar \gamma$, while black denotes temperatures above
that value.

Putting one atom into the cavity (Fig.~\ref{fig:temp1}), we find that
sub-Doppler cooling is achieved in the good-coupling regime $g_0 >
\kappa,\gamma$.  In that regime, the final temperature does not depend
on $g_0$, but is approximately proportional to $\kappa$, as expected
from previous work \cite{domokos01a,domokos02}.  Here one has to be
cautious of the results, since if the atom's kinetic energy is only a
few times the ground-state energy of the harmonic trap potential, the
validity of the semiclassical approximation can be questioned.

Interestingly, the temperature plots look different if we
load two atoms into the cavity (Fig.~\ref{fig:temp2}).  At moderately
large detunings, $\Delta_A=-50\gamma$ and $\Delta_A=-100\gamma$,
temperatures are the same as in the single-atom case. In these cases
the atoms are cooled, but not trapped, by the cavity field.  At larger
detunings, $\Delta_A=-500\gamma$ and larger, we see regions in the
plot where ``extra heating'' compared to the single-atom case is
observed.  This effect is most prominent at extremely large detunings
$\Delta_A=-5000\gamma$ and $\Delta_A=-10^4\gamma$, where for
$\kappa<\gamma$ a new structure appears in the temperature plots.
This points to the highly reduced efficiency of cavity cooling:
whereas for a single atom a better resonator (lower $\kappa$) implies
lower final temperature, if there are two atoms in the cavity,
decreasing $\kappa$ can \emph{increase} the temperature.

Comparison of the temperature plots in Figs.~\ref{fig:temp1} and
\ref{fig:temp2} with the plots of the correlation strength in
Fig.~\ref{fig:corr} reveals strong similarities. Indeed 
one sees that the excess heating caused by the presence of the 
other atom coincides with the buildup of correlations in the motion. 
In other words, the correlation established between two atoms results
in a loss of efficiency of cavity cooling, with final temperatures
pushed up to the Doppler limit.

\subsection{The origin of correlation: numerical tests}
\label{sec:oka_surlodas}
As we have seen above, the motion of the atoms becomes correlated due
to the cavity-mediated cross-talk.  This interaction occurs via the
force, via the friction, and via the diffusion as well.  We would like
to know which of these interaction channels is responsible for the
correlation. In our simulation we can conveniently answer this
question if we artificially weaken only particular channels of
interaction and measure the correlation.

The interaction can be eliminated from the deterministic force using
the approximation \eqref{eq:cavity_pumping_force_large_detuning}.
Friction and diffusion contain interaction at various levels.  On the
one hand, cross-friction and cross-diffusion are direct interactions
between the atoms.  On the other hand -- through the determinant
$\dd'$ -- the positions of both atoms influence the friction and
diffusion constants for the other atoms.  This parametric interaction
was not analyzed, but does not seem to play a prominent role.  Both
cross-friction and cross-diffusion can be eliminated by simply
suppressing the off-diagonal terms in the friction and diffusion
matrices.

\begin{figure}
\begin{center}
\includegraphics[width=7cm]{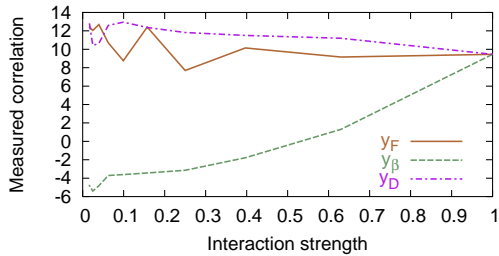}
\caption{\emph{Correlation for various interaction channel
strengths.}
The interaction via the conservative, the friction, and the diffusion
forces is weakened by the mixing ratios $y_F$, $y_\beta$ and 
$y_{\mathcal D}$, as described in the text.
The correlation strength was then measured using our correlation
parameter. 
Cross-friction is seen to play the dominant role in the
establishment of correlation. 
\label{fig:korr_mesterseges} 
}
\end{center}
\end{figure}

The elimination of the interaction can be made continuous with mixing
parameters $0\le y_F,y_\beta,y_\dd \le1$, giving the force, the
friction, and the diffusion with the following formulae:
\begin{align}
  \label{eq:mix_force}
  F' &= y_F F + (1-y_F) F_0,\\
  \beta_{km}' &= y_\beta \beta_{km} 
  + \delta_{km} (1-y_\beta) \beta_{km}; \\
  {\mathcal D}_{km}' &= y_{\mathcal D} {\mathcal D}_{km}
  + \delta_{km} (1-y_{\mathcal D}) {\mathcal D}_{km}.
\end{align}
We show an example of what this gives for detuning 
$\Delta_A=-5000\gamma$, decay rate 
$\kappa=0.5 \gamma$ and coupling $g_0=30\gamma$ in Fig.~\ref{fig:korr_mesterseges}.
The results of these investigations are quite unanimous.
It can be seen that linearization of the potential has no systematic
effect on correlation strength, and correlations are slightly enhanced
if the noise is decorrelated. 
If cross-friction is eliminated, however, all correlations disappear. 
We can therefore conclude that the dominant effect leading to
correlated motion is related to cross-friction. 

\section{The origin of correlation: analytical expansion for deep trapping}
\label{sec:analytical}

The emergence of motional correlations between trapped atoms can be
examined analytically using the formulas presented in Section
\ref{sec:interaction}.  Both friction and diffusion
matrices have an ``isotropic'' term proportional to $\delta_{km}$,
which acts on each atom separately, and an ``interaction'' term,
containing cross-friction and cross-diffusion.  The latter can be
rewritten in the following form, using the notation $g_k =
g(\vect{x}_k)$:
\begin{align}
    \label{eq:beta_es_D_kolcsonhato}
\frac{2 \hbar \eta^2}{\abs{\dd'}^2} 
P(\{\xl\})
\begin{pmatrix}
\vgrad_1 g_1^2\\
\vdots\\
\vgrad_N g_N^2
\end{pmatrix}
\circ
\begin{matrix}
\big( \vgrad_1 g_1^2&
\cdots & 
\vgrad_N g_N^2\big)\\
\vphantom{\vgrad_2 g_2^2} \\
\vphantom{\vgrad_2 g_2^2} \\
\vphantom{\vgrad_2 g_2^2} 
\end{matrix},
\end{align}
where the prefactor $P$ depends on the positions of all atoms
symmetrically. 

For well-trapped atoms $x_k \ll \lambda$ ($x_k$ denoting the distance of
atom $k$ from the nearest trapping site), the coupling constants can
be approximated as
\begin{align}
   \label{eq:g_kicsi}
   g_k = g(x_k) = g_0 \cos (k_C x_k) 
   \approx g_0 \left(1- \frac{1}{2} \, k_C^2 x_k^2\right). 
\end{align}
Substituting these into the interaction parts of the matrices 
\eqref{eq:beta_es_D_kolcsonhato}
yields:
\begin{align}
    \label{eq:beta_es_D_kolcsonhato2}
\frac{8 \hbar \eta^2}{\abs{\dd'}^2} 
P(\{\xl\})
g_0^4 k_C^4 r^2
\begin{pmatrix}
x_1/r\\
\vdots\\
x_N/r
\end{pmatrix}
\circ
\begin{matrix}
\big( x_1/r& 
\cdots & x_N/r\big)\\
\vphantom{x_2)} \\
\vphantom{x_2)} 
\end{matrix},
\end{align}
where the ``radius'', $r=\left({\sum_l x_l^2}\right)^{\frac{1}{2}}$, is the
distance of the system from the origin in the coordinate space.
This matrix is a projector to the vector $\left(x_1, \ldots , x_N
\right)$, and hence  
the interaction terms affect only the ``radial'' motion. 

In the well-trapped case we can write the friction and diffusion
matrices to second order in the coordinates $x_l$ as:
\begin{align}
    \label{eq:form1}
\bkm&={\beta}_0 \delta_{km} + \beta_1 x_k x_m, \\
\dkm&={\mathcal D}_0 \delta_{km} + {\mathcal D}_1 x_k x_m.
\end{align}
The following notation is used:
\begin{align}
    \label{eq:form12}
{\beta}_0 &= 2 \hbar \frac{\eta^2}{\abs{\dd'_0}^2} \gamma g_0^2 k_C^2 
\frac{2\Delta_A}{\Delta_A^2+\gamma^2} k_C^2 x_k^2, \\
{\mathcal D}_0 &= 2 \hbar^2 \frac{\eta^2}{\abs{\dd'_0}^2} \gamma g_0^2 k_C^2 
\left( \frac{2}{5} + \frac{3}{5} k_C^2 x_k^2\right), \\
{\beta}_1 &= 2 \hbar \frac{\eta^2}{\abs{\dd'_0}^2} g_0^4 k_C^4
   \Im \left\{ \frac{1}{{\dd'_0}^2} \biggl[ \dd'_0 (1-3\chi)  + \right. \\
   &\left. \biggl. 2(1-\chi)  \Bigl( (i\Delta_A-\gamma)^2 - \sum_l g(\xl)^2 \Bigr)   
       \biggr] \right\}, \\
{\mathcal D}_1 &= 2 \hbar^2 \frac{\eta^2}{\abs{\dd'_0}^2} g_0^4 k_C^4\,
4\frac{\kappa \Delta_A +\gamma \Delta_C}{\abs{\dd'_0}^2},
\end{align}
where $\dd'_0$ stands for the Bloch determinant $\dd'$ defined in 
eq.~(\ref{d_prime_def}), evaluated at the trapping site.
Near the trapping site all $x_k$ go to zero, and only $\dd_0$ remains
finite. 
This prevents the atoms from stopping completely: they are ``heated
out'' from the trapping sites themselves. 
At some distance from the
trapping sites the interaction terms become important as well, in the
coordinate space this induces extra diffusion and friction 
in the radial direction. 
Provided $g_0$ is high enough, radial friction is enhanced much more
than radial diffusion, leading to a ``freezing out'' of the radial
mode, i.e.~a motion at approximately constant distance from the origin. 

\section{Conclusions}  

As pointed out in several previous papers, the motion of particles in a
cavity is coupled quite generally through the field mode. This implies
the buildup of motional correlations, which we have investigated here in
more detail. In general we found that in steady state these
correlations are hard to see directly and it is difficult to find good
qualitative measures to characterize them. Mostly they are strongly
perturbed by various diffusion mechanisms. However, they still play
an important role in the dynamics and thus can be observed indirectly. As one
consequence they can lead to a significant change (increase) in the
steady temperature, which directly relates to trapping times and
localization properties. This poses limits to cavity induced cooling
for large particle numbers.

From the various mechanisms at work, cross-friction turns out to be
the most important.  It creates correlation and leads to a fast
thermalization of two distant particles without direct interaction.
Hence, this mechanism should prove vital for the implementation of
sympathetic cooling of distant ensembles coupled by a far off-resonant
cavity field. In order to get efficient coupling the two species
should have comparable oscillation frequencies, so that correlation
buildup and thermalization is fast. The finesse of the cavity should
also be large (long photon lifetime).
 
The second important coupling mechanism, which works via the joint
steady state potentials, was suggested for use in the implementation
of conditional phase shifts \cite{hemmerich99}. It can be viewed as a
cavity-enhanced dipole--dipole coupling.  Although this contribution
will become more important for larger atom-field detunings, the
conditions for this part to dominate the dissipative cross-friction
seem rather hard to achieve in practice.  Finally we found that the
noise forces acting on different atoms contain nonlocal correlations
too. These are particularly important for large detunings and
relatively low photon numbers, where spontaneous emission is strongly
reduced. As a result they could seriously perturb bipartite quantum
gates in cavities.
   
\section*{Acknowledgments}
We would like to thank A.~Vukics for helpful discussions. This work was
supported by the National Scientific Research Fund of Hungary (OTKA)
under contracts Nos.~T043079 and T034484 and through Project 12 of SFB
Quantenoptik in Innsbruck of the Austrian FWF.
P.~D.~ackowledges support by the Hungarian Academy of Sciences
(Bolyai Programme).

\end{document}